\documentclass[12pt]{extarticle}

\usepackage[sorting=none,style=numeric-comp,giveninits=true]{biblatex}
\usepackage{enumitem}
\RequirePackage[top=20.2mm,bottom=20.2mm,left=20.2mm,right=20.2mm]{geometry}

\usepackage{fancyhdr}
\usepackage{mathtools}
\usepackage{subfigure}
\usepackage{amssymb}
\usepackage{amsmath}
\usepackage{hyperref}
\usepackage{amsfonts}
\usepackage{amssymb}
\usepackage{caption}
\usepackage{float}
\usepackage{bm}
\usepackage{graphicx}
\usepackage{color}
\usepackage{verbatim}
\usepackage{blindtext, multicol}
\usepackage{multirow}
\usepackage{epigraph}
\usepackage[dvipsnames]{xcolor}
\usepackage{physics}

\usepackage{cancel}
\usepackage{etoolbox}

\usepackage[english]{babel}
\usepackage{csquotes}
\DeclareUnicodeCharacter{0301}{\'{i}}
\usepackage[nottoc]{tocbibind}

\usepackage{academicons}
\usepackage{xcolor}
\usepackage{fontawesome5}
\definecolor{orcidlogocol}{rgb}{0.65, 0.807, 0.223}

\newcommand{\orcid}[1]{$\,$\href{https://orcid.org/#1}{\textcolor{orcidlogocol}{\faOrcid}}}
\def \blue {\color{blue}}
\usepackage[T1]{fontenc} 

\usepackage[table]{xcolor}
\usepackage{array}
\usepackage{colortbl}

\definecolor{tablegray}{gray}{0.55}

\def\beq{\begin{equation}}
\def\eeq{\end{equation}}
\def\ber{\begin{eqnarray}}
\def\eer{\end{eqnarray}}
\def\benu{\begin{enumerate}}
\def\eenu{\end{enumerate}}

\def\d{\mathrm{d}}
\def\l{\left}
\def\r{\right}
\def\f{\frac}
\def\mpl{m_{\rm p}}

\def \blue {\color{blue}}
\def \Blue {\color{Blue}}

\def \Brown {\color{Brown}}

\def\SSMR#1{{\color{red!15!black} #1}} 
\def\SSMB#1{{\color{blue!90!black} #1}}

\def \lleq {\lower0.9ex\hbox{ $\buildrel < \over \sim$} ~}
\def \ggeq {\lower0.9ex\hbox{ $\buildrel > \over \sim$} ~}

\hypersetup{
    colorlinks = true,
    linkcolor = Blue,
    citecolor = blue,
    urlcolor = Brown
        }

\usepackage{authblk}
\usepackage{framed}
\colorlet{shadecolor}{CadetBlue!5}

\begin{document}

\title{\textbf{\SSMR{\hspace{-0.3in}Effective Phantom Dark Energy: What Cosmological Reconstruction Does and Does Not Imply}}\vspace{0.1in}}
\author[a]{{\bf {\Blue Swagat S. Mishra}} \orcid{0000-0003-4057-145X}}
\affil[a]{{\small Cosmology, Gravity, and Astroparticle Physics Group, Center for Theoretical Physics of the Universe (CTPU-CGA), Institute for Basic Science (IBS), Daejeon, 34126, Korea}\\

\vspace{0.2in}

{\small Email ID: \texttt{\blue swagatmishra@ibs.re.kr}}
}

\date{}

\maketitle

\begin{abstract}
In observational cosmology, the dark energy density and equation of state are effective quantities reconstructed at the background level under a set of assumptions. These include the FLRW framework, the standard Friedmann equation of General Relativity, and separately conserved non-relativistic matter at late times. Recent analyses involving DESI BAO measurements combined with CMB and supernova data have shown mild preference for dynamical dark energy featuring phantom or phantom-crossing behaviour. While the statistical significance of these trends remains limited, and unresolved systematics or modelling uncertainties may still be important, the resulting discussions have highlighted the need for a clearer interpretation of effective dark energy reconstruction. In particular, effective phantom behaviour does not necessarily imply the existence of a fundamental phantom field, microscopic ghost instabilities, violation of the null energy condition by the fundamental stress tensor, or a catastrophic cosmic future. The purpose of this work is to clarify these distinctions, independently of whether the current observational preference for dynamical dark energy survives future data. We discuss the definition of effective dark energy in cosmology, the interpretation of phantom and phantom-crossing behaviour, introduce a simple kinematic criterion for identifying effective phantom evolution directly from the expansion history, and review physical mechanisms through which effective phantom behaviour may arise without fundamental pathologies. While familiar within the dark energy reconstruction community, these distinctions are often left implicit in broader discussions of dynamical dark energy. We hope that this work will remain useful beyond the present observational situation as a clarification of what observationally reconstructed dark energy does and does not imply.
\end{abstract}

\vspace{0.2in}


\newpage

\begin{small}
\tableofcontents
\end{small}


\medskip

\begin{shaded}
\vspace{-0.25in}
\begin{center}
\subsection*{\underline{\Brown Units, Notation, and Convention}}
\end{center}
\begin{itemize}
\item Throughout this work, we consistently refer to the present-epoch value of any cosmological observable (say ${\cal F}$) with a subscript `$0$', namely, ${\cal F}(z=0) \equiv {\cal F}_0$. 
\item  We work with FLRW metric, with scale factor $a$, and  the corresponding redshift $1+z = a_0/a$. For simplicity, we only discuss the case of spatially flat FLRW Universe with $\Omega_K = 0$.
\item We mostly use natural units $\hbar=c=1$  (unless explicitly specified), and the reduced Planck mass is $\mpl=1/\sqrt{8\pi G}$.
\item Hubble parameter is denoted by $H(z)$, and its dimensionless version by $\boxed{{\cal H}(z)\equiv \f{H(z)}{H_0}}$.
\item Dimensionless cosmological density parameters are written as $\Omega_{{\cal F}}(z)$, whereas $\Omega_{{\cal F}0}$ denotes its corresponding value at the present epoch. These are explicitly given by
\beq
\boxed{\Omega_{{\cal F}}(z) \equiv  \f{\rho_{\cal F}(z)}{3\,\mpl^2\,H^2(z)}} \qquad \Longrightarrow \qquad \boxed{\Omega_{{\cal F}0} \equiv  \f{\rho_{{\cal F}0}}{3\,\mpl^2\,H_0^2}}\, ,
\nonumber
\eeq
where, the critical densities are given by:~~ $\boxed{\rho_c(z) \equiv 3\,\mpl^2\,H^2(z)} ~ \Longrightarrow ~ \boxed{\rho_{c0} \equiv 3\,\mpl^2\,H_0^2}\, .$\\

Note that, while in most observational cosmology papers, $\Omega_{{\cal F}0}$ is denoted as $\Omega_{\cal F}$ (implying the present-epoch density), we avoid such convention as it can potentially lead to confusion.  
\item The effective  density and equation-of-state parameter of dark energy, as reconstructed from the background expansion, are denoted by $\rho_{\rm DE}^{\rm eff}$ and $w_{\rm DE}^{\rm eff}$.
\item The `equation-of-state parameters', $\lbrace w_{\cal F},\,w_{\rm DE}^{\rm eff},\,w_{\rm tot},...\rbrace$, will simply be referred to as `equations of state', following standard cosmological terminology.
\item Derivatives with respect to cosmic time are denoted by an overdot, while derivatives with respect to redshift are written explicitly as $\f{\d}{\d z}$.
\end{itemize}
\end{shaded}
\newpage

\section{Introduction}
\label{sec:Intro}
The discovery of late time accelerated expansion of the Universe~\cite{Ostriker:1995su,SupernovaSearchTeam:1998fmf,SupernovaCosmologyProject:1998vns} established dark energy (DE) as one of the central problems of modern cosmology~\cite{Sahni_2004,Amendola:2015ksp}. In the simplest description, cosmic acceleration is sourced by an effective cosmological constant, whose equation of state (EoS) satisfies $w_\Lambda=-1$, or equivalently, $p_\Lambda = -\rho_\Lambda$. This description remains remarkably successful phenomenologically~\cite{Hinshaw_2013,Planck:2018nkj,AtacamaCosmologyTelescope:2025blo,SPT-3G:2025bzu}. At the same time, the smallness of the inferred vacuum energy~\cite{Zeldovich:1968ehl,Weinberg:1988cp,Sahni:1999gb,Padmanabhan_2003,Bousso_2007,Padilla:2015aaa}, the coincidence problem, and the absence of a compelling microscopic explanation motivate the continued exploration of dynamical dark energy, modified gravity, and more general effective descriptions of the late-time Universe~\cite{Ratra:1987rm,Frieman:1995pm,Peebles:2002gy,Peebles:1987ek,Copeland:2006wr, SAHNI_2006,Frieman_2008,durrer2008darkenergymodifiedgravity,Li_2011,Nojiri_2011,Clifton_2012,Caldwell:2005tm,Bamba:2012cp}.

Recent observations have made this question especially timely. In particular, DESI DR2 Baryon Acoustic Oscillation (BAO) measurements~\cite{DESI:2024mwx,DESI:2024aqx,DESI:2025zgx}, when combined with Cosmic Microwave Background (CMB) and Type Ia supernova data~\cite{Planck:2018nkj,2020Planck,AtacamaCosmologyTelescope:2025blo,SPT-3G:2025bzu,Rubin:2023ovl,Brout:2022vxf,DES:2024tys}, seem to suggest a mild preference for dynamical dark energy over $\Lambda$CDM in commonly used extensions of the standard cosmological model~\cite{Wolf:2025jlc,Tada:2024znt,Shlivko:2024llw,Payeur:2024dnq,Keeley:2025stf,Bhattacharya:2024hep,Jiang:2024xnu,Cline:2025sbt,Gialamas:2025pwv}. In the CPL $\lbrace w_0,\,w_a\rbrace$ parametrisation~\cite{Chevallier:2000qy,Linder:2002et}, the preferred region lies in a part of parameter space where the reconstructed dark energy equation of state can cross the phantom divide ($w_{\rm DE} = -1$), being phantom like ($w_{\rm DE} < -1$) in part of the recent past while remaining non-phantom ($w_{\rm DE} > -1$) closer to  the present epoch. The DESI extended dark energy analysis~\cite{DESI:2025fii} further reports that similar qualitative trends appear across several parametric and non-parametric approaches, and that models featuring phantom crossing are preferred, although alternatives without such crossing cannot yet be ruled out. At the same time, the statistical significance of the current preference is not decisive, and several issues remain under discussion, including possible systematics, dataset consistency, parametrisation dependence, prior dependence, and the interpretation of model comparison under Bayesian and frequentist criteria~\cite{Popovic:2025glk,DES:2025sig,cortes2025desisdr2exclusionlambdacdm,Efstathiou:2025tie,Gialamas:2024lyw,Efstathiou:2024xcq,Herold:2025hkb,Ong:2025utx,Ong:2026tta,Keeley:2025rlg,Woo:2026ice,Colgain:2025nzf,Das:2026hfp}. Thus, the observational status of dynamical dark energy  remains unsettled~\cite{CosmoVerseNetwork:2025alb}.

Independent of whether the present observational preference survives further scrutiny and future data, the appearance of such phantom like behaviour~\cite{Caldwell:1999ew} of DE naturally raises theoretical concerns. A fundamental DE component with an EoS $w<-1$ and positive energy density $\rho > 0$ violates standard energy condition expectations~\cite{Dubovsky:2005xd,Buniy:2006xf,Buniy:2005vh,Fewster:2002ne,Ford:2003qt,Elder:2013gya} and is often associated with wrong-sign  kinetic energy, ghost instabilities, and vacuum decay~\cite{Caldwell:2003vq,Nojiri:2003ag,Singh:2003vx,Dabrowski:2003jm,Johri:2003rh,Gonzalez-Diaz:2003xmx,Sami:2003xv,Chimento:2003qy,Barrow:2004xh,Ludwick:2017tox}. If such behaviour persists into the future, the dark energy density can grow without bound,  leading to a  Big Rip singularity in the finite future~\cite{Chimento:2004ps,Copeland:2006wr}. These concerns are not artificial; they are well motivated from the point of view of fundamental theory. They explain why claims of phantom dark energy are often viewed with caution by particle physicists, gravitational theorists, and cosmologists interested in ultraviolet completion or stability~\cite{Sbisa:2014pzo,Caldwell:2003vq,Carroll:2003st,Koyama:2007za,Woodard:2006nt,Cline:2003gs,Garriga:2012pk,Vikman:2004dc}.

However, an important part of the discussion concerns the meaning of the words `{\em dark energy density}', denoted by $\rho_{\rm DE}(z)$,  and its `{\em equation of state}', denoted as $w_{\rm DE}(z)$, in observational cosmology. Cosmological observations do not directly probe the microscopic theory underlying dark energy. Instead, dark energy is reconstructed as an effective residual fluid after assuming a specific cosmological background framework: FLRW symmetry, the standard Friedmann equation of General Relativity, and separately conserved pressureless (dark and baryonic)  matter at late times. Thus, $\rho_{\rm DE}(z)$ and $w_{\rm DE}(z)$ are not fundamental observables in the same sense as a distance, redshift, or clustering scale. It is a {\em reconstructed descriptor} of the {\em residual expansion history} under stated assumptions. This distinction is often clear in technical reconstruction work, but it is not always explicit in broader discussions of dynamical dark energy in cosmology.

This distinction is especially important for a potentially phantom like behaviour of dark energy. In particular, an effective phantom behaviour need not imply the existence of a fundamental field with a kinetic energy term that has wrong sign  in the Lagrangian. It also need not necessarily imply violation of the null energy condition by the underlying microscopic stress tensor, nor does it guarantee a future catastrophic singularity~\cite{Fang:2008sn,Hu:2004kh,Caldwell:2005ai}. It is first and foremost a statement about the behaviour of an effective component defined by subtracting an assumed matter contribution from the observed expansion history, under the aforementioned assumptions. 

 There are several familiar mechanisms through which such effective phantom behaviour can
arise without incorporating fundamental phantom matter. In modified gravity, rewriting the background evolution in the form of a standard Friedmann equation forces geometric modifications into an {\em effective dark energy sector}~\cite{Sahni:2002dx,Clifton:2011jh,Joyce:2016vqv}. In interacting dark sector models~\cite{Amendola:1999er,Hu:2004kh}, the assumption of separately conserved pressureless matter need not hold, and interpreting the same expansion history using a conserved matter template, with $\rho_m \propto a^{-3}$, can produce an anomalous effective equation of state of dark energy. On the other hand, in scalar-tensor theories, $f(R)$ models, and braneworld cosmologies, the effective four dimensional dark energy sector can satisfy $w_{\rm DE}^{\rm eff} < -1$ even when the underlying degrees of freedom are not fundamental phantom ghosts~\cite{Esposito-Farese:2000pbo,Sahni:2002dx,Amendola:2007nt,Gannouji:2006jm,Sahni:2014ooa,Bag:2018jle,Mishra:2025goj,Alam:2016wpf}. 
These examples are not invoked here as preferred explanations of the current datasets. Their role is simply to demonstrate that effective phantom behaviour and microscopic phantom dynamics are logically distinct.  

\vspace{-0.1in}
\begin{shaded}
\noindent{\bf \Brown Main purpose of this work:}~~Following the aforementioned discussions, the primary purpose of this work is to clarify the meaning and interpretation of effective dark energy reconstruction in observational cosmology, and to place recent discussions of phantom and phantom crossing dark energy in their proper conceptual context. We do not attempt to assess the statistical robustness of the current DESI preference, nor do we advocate a particular model of dynamical dark energy. The scope of this note is therefore deliberately limited. The point is instead conceptual:
\begin{enumerate}
\item To state explicitly what is assumed when DE is reconstructed in observational cosmology. 
\item What `{\em effective phantom} and {\em phantom-crossing behaviour}' mean within that reconstruction. 
\item And finally, what conclusions do or do not follow from such behaviour.
\end{enumerate}

In hindsight, such a clarification could have been written independently of the latest observational developments. Recent discussions have nevertheless made the issue timely. In the author's experience, conversations across the  cosmology, gravitational theory, and high energy physics  communities often involve different implicit assumptions about what reconstructed dark energy quantities represent. This note is intended to make those assumptions explicit and to separate effective reconstruction from microscopic interpretation.
\end{shaded}

Many of the conceptual distinctions emphasized in this note are not completely new, and are well known within the dark energy reconstruction community~\cite{Sahni:2006pa,Saini:1999ba,Boisseau:2000pr,Huterer:1998qv,Kujat:2001ke,Sahni:2002fz,Gerke:2002sx,Alam:2003sc,Alam:2003fg,Copeland:2006wr,Sahni:2008xx,Shafieloo:2012rs,Weinberg_2013}. In particular, the earlier dark energy literature from the early 2000s already discussed several scenarios in which an effective phantom or phantom crossing behaviour could arise without introducing a fundamental phantom field, microscopic ghost instabilities, or a future Big Rip singularity~\cite{Fang:2008sn,Hu:2004kh,Caldwell:2005ai,Sahni:2002dx,Clifton:2011jh,Joyce:2016vqv,Amendola:1999er,Hu:2004kh,Amendola:2007nt,Esposito-Farese:2000pbo,Kunz:2006wc,Perivolaropoulos:2005yv,Arefeva:2005mka,Vikman:2004dc,Elizalde:2004mq,Nojiri:2003vn,Nojiri:2005sr}. These developments were partly motivated by early supernova based dark energy reconstruction and parametrization studies which mildly hinted towards evolving, phantom like, or phantom crossing dark energy~\cite{SupernovaSearchTeam:2004lze,Alam:2003fg,Riess:2006fw}. However, following the observational success of $\Lambda$CDM over the past decade and half, many of these distinctions have gradually become less explicit in broader discussions of dark energy. The recent DESI results have once again brought these issues to the forefront, making it timely to revisit and systematically clarify the interpretation of effective phantom behaviour in cosmology. We also derive a simple kinematic characterization of effective phantom and phantom-crossing dark energy directly in terms of the Hubble expansion history, leading to a quantity that we refer to as the {\em phantom-crossing marker}.

Throughout this work, our discussion is deliberately restricted to the reconstruction of dark energy at the homogeneous background level through the cosmological expansion history $H(z)$. Inclusion of cosmological perturbations provides additional information regarding the underlying microscopic dynamics through quantities such as structure growth, gravitational potentials, clustering properties, and effective gravitational couplings~\cite{Copeland:2006wr,Weinberg_2013,Huterer:2017buf}. Different microscopic theories may therefore produce similar background reconstructions while differing substantially at the perturbative level~\cite{Kunz:2007rk,Joyce:2016vqv,Bag:2018jle}. Nevertheless, the central conceptual distinction emphasized in this work between observationally reconstructed effective dark energy and the underlying microscopic theory already arises at the level of background cosmology itself.

\bigskip

The rest of the paper is organized as follows. Sec.~\ref{sec:Eff_DE}  reviews the assumptions entering the cosmological reconstruction of effective dark energy and provides definitions of the effective  density and equation of state of dark energy. Sec.~\ref{sec:DDE_DESI_Obs} discusses the recent observational hints of dynamical dark energy, with emphasis on what DESI results imply and what they do not imply. Sec.~\ref{sec:Eff_DE_Pheno} briefly reviews mechanisms through which effective phantom behaviour can arise without fundamental instabilities, including a braneworld illustration. Sec.~\ref{sec:Summary} summarizes the main conclusions.

\section{Cosmological Reconstruction of Dark Energy}
\label{sec:Eff_DE}

In this section, which constitutes the core of this paper, we carefully lay out the assumptions, definitions, and notation underlying the cosmological reconstruction of dark energy. Since much of the confusion surrounding recent observational hints of dynamical (including  phantom-crossing) dark energy originates from the interpretation of reconstructed effective quantities, it is important to clearly distinguish between directly observed cosmological quantities and inferred effective descriptions. Particular attention should therefore be paid to the assumptions entering the {\em  definitions of the effective density} and {\em equation of state of dark energy}.

\subsection{Effective Dark Energy in Cosmology: Assumptions and Definitions}
\label{sec:Eff_DE_Assumptions}
 At sufficiently low redshift $z\lesssim 10$, where the radiation contribution to the background expansion can be neglected, the total cosmological energy density is well described by non-relativistic (dark and baryonic) matter and the component conventionally called {\em dark energy}.  In observational cosmology, from the perspective of reconstructing the properties of dark energy, expansion of the Universe is typically described by
\beq
\boxed{~H^2(z) \equiv \f{1}{3\,\mpl^2} \, \rho_{\rm tot}(z) = \f{1}{3\,\mpl^2}\biggl[\rho_{m0}\l(1+z\r)^3 + \rho_{\rm DE}^\text{eff}(z) \biggr]~} \, .
\label{eq:Friedmann_GR_eff}
\eeq
The {\em effective energy density} of  dark energy is then defined as the residual quantity
\beq
\boxed{~\bm{\rho_{\rm DE}^\text{eff}(z) = 3\,\mpl^2\,H^2(z) - \rho_{m0}\,\l(1+z\r)^3} ~} \, .
\label{eq:rho_DE_eff_def}
\eeq
Several important remarks should immediately be made regarding Eqs.~\eqref{eq:Friedmann_GR_eff}~and~\eqref{eq:rho_DE_eff_def}. First, the quantity $\rho_{\rm DE}^{\rm eff}(z)$ is not directly observed. Cosmological observations constrain quantities such as luminosity distances, angular diameter distances, baryon acoustic oscillation scales, and the expansion history $H(z)$. The quantity $\rho_{\rm DE}^{\rm eff}(z)$ is instead reconstructed after assuming a specific decomposition of the background expansion into {\em pressureless matter} plus a {\em residual dark energy}  sector. Second, this reconstruction implicitly assumes a specific cosmological framework. In particular, Eq.~\eqref{eq:Friedmann_GR_eff} assumes:
\begin{shaded}
\vspace{-0.1in}
\begin{center}
\underline{\bf \Brown Assumptions in Dark Energy Reconstruction}
\end{center}
\begin{enumerate}
\item FLRW symmetry at the background level.
\item The standard Friedmann equation, Eq.~\eqref{eq:Friedmann_GR_eff}, of General Relativity.
\item A pressureless non-relativistic matter component satisfying
\beq
\boxed{~\bm{\rho_m(z)=\rho_{m0}\,(1+z)^3}~} \, .
\label{eq:matter_scaling}
\eeq
\item Separate conservation of this pressureless matter sector and the residual dark energy sector. Since this effective dark energy component is assumed to be separately  conserved, with its pressure and density being related by $p_{\rm DE}^{\rm eff}(z) = w_{\rm DE}^{\rm eff}(z) \rho_{\rm DE}^{\rm eff}(z)$, and
\beq
\f{\d}{\d t}\,\rho_{\rm DE}^{\rm eff}(z)
=
-3H\rho_{\rm DE}^{\rm eff}(z)
\Bigl(1+w_{\rm DE}^{\rm eff}(z)\Bigr) \, ,
\label{eq:DE_continuity}
\eeq
one obtains the {\em effective equation of state of dark energy} as
\beq
\boxed{\bm{w_{\rm DE}^{\rm eff}(z) \equiv -1 + \l(\f{1+z}{3}\r)\, \f{\d}{\d z} \, {\rm ln} \,\rho_{\rm DE}^{\rm eff}(z) }}\, .
\label{eq:w_eff_logrho} 
\eeq
\end{enumerate}
\begin{center}
\underline{\bf \Brown Effective Interpretation of the Reconstructed Dark Energy}
\end{center}

Under the aforementioned assumptions, the difference between the reconstructed expansion history, represented by $3\mpl^2H^2(z)$, and the assumed pressureless matter contribution, $\rho_m(z)\propto (1+z)^3$, is by definition assigned to the effective dark energy sector $\rho_{\rm DE}^{\rm eff}(z)$. A key conceptual point is that although $\rho_m(z)$ represents the matter sector of the Universe, within the reconstruction framework discussed above its role is effectively that of a reference background component with the standard conserved scaling $\rho_m(z)\propto (1+z)^3$, following Eq.~\eqref{eq:matter_scaling}. In this sense, the reconstruction of effective dark energy is fundamentally a background-level or kinematic procedure. \\

It is important to emphasize that Eq.~\eqref{eq:rho_DE_eff_def} remains mathematically meaningful even when the underlying theory does not contain a fundamental dark energy fluid. For instance, modified gravity, interacting dark sector models, braneworld cosmology, or departures from standard matter evolution can all modify the structures of Eqs.~\eqref{eq:Friedmann_GR_eff}~and/or~\eqref{eq:matter_scaling}. However, when such scenarios are rewritten in the form of Eqs.~\eqref{eq:Friedmann_GR_eff}~and~\eqref{eq:matter_scaling}, the modifications are absorbed into an effective residual component which is then {\em interpreted as dark energy}. \\

$\Longrightarrow$~\fbox{
\begin{minipage}{0.85\textwidth}
\noindent Thus, $\rho_{\rm DE}^{\rm eff}(z)$ should primarily be viewed as a  {\em reconstructed descriptor of the residual background expansion history}.\end{minipage}}
\end{shaded}
\noindent In reference to Eq.~\eqref{eq:matter_scaling}, note that  significant departures from the standard evolution of the matter sector are already strongly constrained by CMB observations and their combination with other cosmological datasets. Nevertheless, from the perspective of dark energy reconstruction, any residual departure from the assumed matter evolution would be absorbed into the effective dark energy sector~\cite{Wang:2016lxa}.

The redshift evolution of the effective density of dark energy, Eq.~\eqref{eq:rho_DE_eff_def},  can be written at the level of the Hubble expansion history (or, background kinematics), using  Eq.~\eqref{eq:matter_scaling}, as 
\beq
\boxed{~
\bm{\f{\d \rho_{\rm DE}^{\rm eff}}{\d z}  =  6\,\mpl^2\,H^2\,\l[\f{\d \, {\rm ln}\,H}{\d z}  - \f{3}{2}\, \f{\Omega_m(z)}{1+z}\r]  = 6\,\mpl^2\,H_0^2\,\l[{\cal H}\f{\d\, {\cal H}}{\d z} - \f{3}{2}\, \Omega_{m0}\,(1+z)^2\r]}~}\, .
\label{eq:DE_Density_Evolution}
 \eeq 
Furthermore, in terms of the present-epoch {\em normalized effective dark energy density}
\beq
f_{\rm DE}(z)\equiv\f{\rho_{\rm DE}^{\rm eff}(z)}{\rho_{{\rm DE}0}^{\rm eff}} \, ,
\label{eq:fde_def}
\eeq
the effective EoS can be written as
\beq
\boxed{~w_{\rm DE}^{\rm eff}(z)= -1 + \l(\f{1+z}{3}\r)\, \f{\d}{\d z} \, {\rm ln} \,f_{\rm DE}(z)~}\,.
\label{eq:w_fde}
\eeq
For completeness, we mention that the effective EoS of dark energy can also be written in terms of purely kinematic background quantities. Defining the deceleration parameter (remember, ${\cal H}(z)\equiv H(z)/H_0$ in our notation)
\beq
q(z)\equiv -\f{\ddot a}{a\,H^2}= \f{1+z}{{\cal H}(z)} \, \f{\d}{\d z}\,{\cal H}(z)-1\, ,
\label{eq:q_def}
\eeq
 together with
\beq
\Omega_m(z)=\f{\rho_{m0}\,(1+z)^3}{3\,\mpl^2\,H^2(z)}=\Omega_{0m}\,\f{(1+z)^3}{{\cal H}^2(z)} \, ,
\label{eq:Omega_m_z}
\eeq
one finds
\beq
\boxed{\bm{w_{\rm DE}^{\rm eff}(z)=\f{2\,q(z)-1}{3\Bigl[1-\Omega_m(z)\Bigr]}}}\,.
\label{eq:w_q}
\eeq
Eq.~\eqref{eq:w_q} makes explicit that the {\em reconstructed effective equation of state is fundamentally a kinematic characterization of the background expansion history}, with  $\Omega_{0m}$ being a free parameter of the model.

\medskip

It is useful to distinguish the effective EoS of dark energy, $w_{\rm DE}^{\rm eff}$,  from the (total) {\em effective equation  of state of the Universe}, which  is directly related to the deceleration parameter as
\beq
\boxed{~w_{\rm tot}(z) \equiv \f{p_{\rm tot}(z)}{\rho_{\rm tot}(z)} = -\f{1}{3} + \f{2}{3}\, q(z)~}\, ,
\label{eq:w_tot_def}
\eeq
such that accelerated expansion corresponds to $q < 0~\Rightarrow~w_{\rm tot}< -1/3$. 
Equivalently, for pressureless matter plus the reconstructed effective dark energy component,
\beq
\boxed{w_{\rm tot}(z) = w_{\rm DE}^{\rm eff}(z) \, \Omega_{\rm DE}^{\rm eff}(z) = w_{\rm DE}^{\rm eff}(z)\,\l[1-\Omega_m(z)\r]}\, ,
\label{eq:w_tot}
\eeq
where
\beq
\Omega_{\rm DE}^{\rm eff}(z)=\f{\rho_{\rm DE}^{\rm eff}(z)}{\rho_m(z)+\rho_{\rm DE}^{\rm eff}(z)}\,.
\label{eq:Omega_DE_eff}
\eeq
Therefore, $w_{\rm DE}^{\rm eff}(z)$ and $w_{\rm tot}(z)$ are distinct quantities. In particular, an effective dark energy sector satisfying $w_{\rm DE}^{\rm eff}<-1$ does not necessarily imply $w_{\rm tot}<-1$. This crucial distinction will be important later when discussing the null energy condition in Sec.~\ref{sec:DESI_DR2_Implications}.

\begin{shaded}
\noindent{\bf \Brown Important:}~Eq.~\eqref{eq:w_eff_logrho} is the basic reconstruction equation used throughout observational analysis  of dark energy. It makes explicit that the reconstructed equation of state is determined by the redshift evolution of the residual component defined in Eq.~\eqref{eq:rho_DE_eff_def}. In particular, the quantity $w_{\rm DE}^{\rm eff}(z)$ is not a primary cosmological observable. Rather, it is inferred only after:~\textit{(i)}~reconstructing the expansion history,~\textit{(ii)}~specifying the matter contribution,~\textit{(iii)}~ defining the residual dark energy sector, and~\textit{(iv)}~assuming separate conservation of that residual component. Schematically, the reconstruction proceeds as
\beq
{\rm\textbf{Observables}}~\Longrightarrow~{\rm\textbf{H(z)}}~\Longrightarrow~\bm{\rho_{\rm DE}^{\rm eff}(z)}~\Longrightarrow~\bm{w_{\rm DE}^{\rm eff}(z)}~\Longrightarrow~{\rm\textbf{Model Interpretation}}\,.
\label{eq:reconstruction_hierarchy}
\eeq
The distinction between these different levels is conceptually important. The directly measured quantities are distances, clustering scales, and related observables. The inferred quantity $w_{\rm DE}^{\rm eff}(z)$ already depends on assumptions regarding the cosmological background evolution, Eq.~\eqref{eq:Friedmann_GR_eff}, and the matter sector, Eqs.~\eqref{eq:matter_scaling}~and~\eqref{eq:DE_continuity}. Consequently, different underlying theories may produce very similar reconstructed effective equation of states.
\end{shaded}

Implications of the aforementioned definitions of effective dark energy in the context of the recent observations~\cite{DESI:2024mwx,DESI:2024aqx,DESI:2025zgx,DESI:2025fii} will be discussed in Sec.~\ref{sec:DDE_DESI_Obs}. However, before proceeding to do so, we first discuss the possibility of effective phantom dark energy in Sec.~\ref{sec:DE_Phantom_Eff}, which is conceptually important independent of whether the current observational preference for dynamical dark energy survives further scrutiny.

\subsection{Effective Phantom and Phantom-crossing  Dark Energy}
\label{sec:DE_Phantom_Eff}
A particularly important consequence of the reconstruction equations discussed in Sec.~\ref{sec:Eff_DE_Assumptions} is that the reconstructed dark energy component may exhibit an {\em effective phantom behaviour},
\beq
\boxed{~\bm{w_{\rm DE}^{\rm eff}(z)<-1}~}\,.
\label{eq:wDE_Phantom_Def}
\eeq
Historically, such a behaviour attracted considerable attention since a fundamental phantom field with a wrong-sign kinetic energy can lead to microscopic ghost instabilities and, in some cases, future cosmological singularities such as the Big Rip. However, within the framework of cosmological reconstruction, the interpretation of phantom behaviour is more subtle. Since the reconstructed dark energy density is defined as an effective residual component after assuming a specific background framework, an effective phantom behaviour need not imply the existence of a fundamental phantom degree of freedom. To make this distinction explicit, consider the following\,--

A cosmological scenario\footnote{For example, standard GR with an interacting dark sector and/or modified gravity, as discussed in Sec.~\ref{sec:Eff_DE_Pheno}.} in which the underlying degrees of freedom in the fundamental Action do not contain an inherently  phantom component, and all the fluid constituents ${\cal F}_i$ satisfy 
\beq
w_{{\cal F},i}(z)>-1 \, ,
\label{eq:w_constituents}
\eeq
throughout the redshift range probed by cosmological observations. Then the reconstructed effective dark energy can nevertheless exhibit phantom like behaviour if
\beq
\boxed{~\bm{w_{\rm DE}^{\rm eff}(z)<-1~\Longleftrightarrow~\f{\d \rho_{\rm DE}^{\rm eff}}{\d z}<0~\Longleftrightarrow~\f{\d \rho_{\rm DE}^{\rm eff}}{\d t}>0}~}~~\text{for some}~~z\in(z_{\rm ph},\,z_{\rm max}]\,,
\label{eq:wDE_phantom_equiv}
\eeq
where $z_{\rm ph} \gtrsim 0$ is the lowest redshift up to which the reconstructed dark energy remains effectively phantom. (If $z_{\rm ph} < 0$, the phantom behaviour will be realised in the future.) 

\medskip

The above relation, Eq.~\eqref{eq:wDE_phantom_equiv}, follows directly from Eq.~\eqref{eq:w_eff_logrho}. Thus, effective phantom behaviour corresponds to a reconstructed residual dark energy density that decreases with increasing redshift, or equivalently increases with cosmic time, implying a {\em growing effective dark energy density}. Importantly, this statement is purely at the level of effective dark energy reconstruction. By itself, it neither specifies the microscopic origin of the expansion history nor implies the existence of a fundamental phantom field with wrong-sign kinetic energy or violation of the null energy condition by the underlying stress tensor. Consequently, observational hints of effective phantom behaviour should first be interpreted as a feature of the reconstructed cosmological expansion history under specified assumptions, rather than immediate evidence for a pathological microscopic theory. 

\medskip

An equivalent, and particularly useful, interpretation of effective phantom behaviour may be obtained directly at the level of the Hubble expansion history. From the evolution of the effective density of dark energy, Eq.~\eqref{eq:DE_Density_Evolution}, the effective phantom behaviour, Eq.~\eqref{eq:wDE_Phantom_Def}, can be written as 
\beq
\boxed{~
\bm{ \f{\d \rho_{\rm DE}^{\rm eff}}{\d z}~<~0 \quad \Longleftrightarrow \quad \f{\d \ln H}{\d z}~<~\f{3}{2}\,\f{\Omega_m(z)}{1+z} \quad \Longleftrightarrow \quad {\cal H}(z)\,\f{\d {\cal H}}{\d z} ~<~ \f{3}{2}\,\Omega_{m0}\,(1+z)^2 
}}\, .
\label{eq:Hubble_phantom}
\eeq
Thus, effective phantom dark energy may be interpreted purely as a kinematic property of the
background expansion history. Relative to the assumed conserved matter contribution,
$\rho_m(z)=\rho_{m0}(1+z)^3$, the Hubble parameter evolves with redshift sufficiently slowly
that the residual component defined in Eq.~(2) grows with cosmic time.

\medskip

\begin{shaded}
\noindent{\bf\Brown Crossing the phantom divide}:~~In the context of   {\em phantom-crossing} dynamics of dark energy, the reconstructed effective equation of state evolves as
\beq
\boxed{~\bm{ w_{\rm DE}^{\rm eff}(z) < -1 ~~\l(\text{for}~~z>z_{\rm ph}\r) ~~ \longrightarrow ~~ w_{\rm DE}^{\rm eff}(z) > -1 ~~\l(\text{for}~~z < z_{\rm ph}\r) }~}\,,
\label{eq:wDE_phantom_crossing}
\eeq
thereby crossing the {\em phantom divide} or {\em phantom boundary}~\cite{Fang:2008sn,Hu:2004kh,Caldwell:2005ai,Hu:2004kh,Vikman:2004dc,Kunz:2006wc,Perivolaropoulos:2005yv,Arefeva:2005mka,Amendola:2007nt} at some $z = z_{\rm ph}$.  Since the effective dark energy equation of state is reconstructed from the expansion history, crossing the phantom divide simply corresponds to a change in the sign of the derivative of dark energy density, namely, 
\beq
\boxed{~\bm{\f{\d \rho_{\rm DE}^{\rm eff}}{\d t}  > 0 ~~\l(\text{for}~~z>z_{\rm ph}\r) ~~ \longrightarrow ~~ \f{\d \rho_{\rm DE}^{\rm eff}}{\d t} < 0 ~~\l(\text{for}~~z<z_{\rm ph}\r)}~}\,.
\label{eq:rhoDE_phantom_crossing}
\eeq
\end{shaded}
At the level of effective reconstruction, such a crossing does not by itself imply any singular behaviour or inconsistency in the underlying microscopic theory. Instead, it should first be interpreted, under the assumptions discussed in Sec.~\ref{sec:Eff_DE_Assumptions}, as a scenario in which the residual expansion or effective dark energy density increases with time in the past for $z > z_{\rm ph}$, reaching a maximum at $z=z_{\rm ph}$, before beginning to decrease at lower redshifts. As we discuss later in Sec.~\ref{sec:DDE_DESI_Obs}, this is precisely the qualitative phantom crossing behaviour mildly preferred by the recent DESI reconstruction results~\cite{DESI:2025fii}. Alternatively, the phantom-divide crossing can also be achieved in a situation with reversed chronology, where $\rho_{\rm DE}^{\rm eff}$ decreases with time at higher redshifts, reaches a minimum at $z=z_{\rm ph}$, before beginning to grow at lower redshifts. 

\begin{shaded}
\noindent At the level of the Hubble expansion history, Eq.~\eqref{eq:DE_Density_Evolution}, the {\em phantom boundary} can be written as, 
\beq
\boxed{~\bm{w_{\rm DE}^{\rm eff}(z_{\rm ph}) = -1}} \Longleftrightarrow \boxed{~\bm{\f{\d \ln H}{\d z}\bigg\vert_{z_{\rm ph}}~=~\f{3}{2}\,\f{\Omega_m(z_{\rm ph})}{1+z_{\rm ph}}}~} \Longleftrightarrow  \boxed{~\bm{{\cal H}\,\f{\d {\cal H}}{\d z}\bigg\vert_{z_{\rm ph}} ~=~ \f{3}{2}\,\Omega_{m0}\,(1+z_{\rm ph})^2}} \, ,
\label{eq:Hubble_phantom_crossing}
\eeq
which marks the transition between phantom and non-phantom evolution. Therefore, crossing the phantom divide, Eq.~\eqref{eq:rhoDE_phantom_crossing}, corresponds to
\beq
\boxed{~
\bm{ {\cal H}(z)\,\f{\d {\cal H}}{\d z} ~<~ \f{3}{2}\,\Omega_{m0}\,(1+z)^2  ~~\l(\text{for}~~z>z_{\rm ph}\r) ~~ \longrightarrow ~~ {\cal H}(z)\,\f{\d {\cal H}}{\d z} ~>~ \f{3}{2}\,\Omega_{m0}\,(1+z)^2
}}\, .
\label{eq:Hubble_phantom_Marker}
\eeq
Consequently, the effective phantom behaviour may also be interpreted directly at the level of the background expansion history, as following.  During the phantom phase, the quantity $3\mpl^2H^2(z)$ decreases with time more slowly than the assumed matter contribution $\rho_{m0}(1+z)^3$, which then implies that the residual difference between these two quantities, $\rho_{\rm DE}^{\rm eff}(z)$,  grows with time at redshifts $z > z_{\rm ph}$. Thereafter, this residual component reaches a maximum near $z=z_{\rm ph}$ where it satisfies Eq.~\eqref{eq:Hubble_phantom_crossing}, before beginning to decrease again at lower redshifts for $z< z_{\rm ph}$.
\end{shaded}

An interesting consequence of the above discussion is that crossing of the phantom divide need not occur only once in the cosmological history. Depending upon the evolution of $H(z)$, the quantity $\f{\d\ln H}{\d z}-\f{3}{2}\f{\Omega_m(z)}{1+z}$ may change sign multiple times. Consequently, $w_{\rm DE}^{\rm eff}(z)$ may exhibit multiple phantom crossings, corresponding to alternating epochs of effective phantom ($w_{\rm DE}^{\rm eff}<-1$) and quintessence-like ($w_{\rm DE}^{\rm eff}>-1$) behaviour.

\medskip

Motivated by the above discussion, and under the standard assumptions\,\footnote{The author thanks Shadab Alam for suggesting that the standard assumptions of Sec.~\ref{sec:Eff_DE_Assumptions}  underlying the definition of phantom-crossing marker, especially $\rho_m \propto (1+z)^3$, should be stated explicitly here.} entering the reconstruction of effective dark energy discussed in Sec.~\ref{sec:Eff_DE_Assumptions}, it is useful to define the dimensionless quantity ${\cal C}_{\rm ph}(z)$ as a  {\em phantom-crossing marker}, which provides a simple kinematic criterion for determining whether a reconstructed expansion history corresponds to effective phantom behaviour.
\begin{shaded}
\begin{center}
\vspace{-0.2in}
\underline{\bf \Brown A Phantom-crossing Marker}
\end{center}
The {\em phantom-crossing marker} can be suitably defined as
\beq
\boxed{~\bm{{\cal C}_{\rm ph}(z) ~ \equiv ~ \f{2\,{\cal H}(z)\,\f{\d {\cal H}(z)}{\d z}}{3\,\Omega_{m0}\,(1+z)^2} ~ = ~ \f{2 \, (1+z) \, \f{\d \ln H (z)}{\d z}}{3\,\Omega_{m}(z)}~}} \, ,
\label{eq:Phantom_Diagnostic}
\eeq
such that\,--
\begin{itemize}
\item  A phantom-like behaviour of the effective dark energy corresponds to ~$
\bm{{\cal C}_{\rm ph}(z) < 1}$\, 
\item While, a non-phantom or quintessence-like behaviour corresponds to
~$\bm{{\cal C}_{\rm ph}(z) > 1}$.
\item Hence, the phantom boundary is marked by ~$\bm{{\cal C}_{\rm ph}(z_{\rm ph}) = 1}$.
\end{itemize}

Therefore, if one is given an expansion history $H(z)$, either from cosmological observations or from a theoretical model, the phantom-crossing marker ${\cal C}_{\rm ph}(z)$ provides a direct way of determining whether the corresponding reconstructed dark energy is phantom-like, quintessence-like, or exhibits phantom-crossing behaviour. In this sense, ${\cal C}_{\rm ph}(z)$ may be viewed as a purely kinematic characterisation of the reconstructed dark energy sector, requiring only the background expansion history and the assumption of a conserved pressureless matter component, Eq.~\eqref{eq:matter_scaling}, as emphasised in Sec.~\ref{sec:Eff_DE_Assumptions}.
\end{shaded}
 In Sec.~\ref{sec:Eff_DE_Braneworld}, we will explicitly track the evolution of ${\cal C}_{\rm ph}(z)$ in a braneworld model and show how the transition ${\cal C}_{\rm ph}<1 \rightarrow {\cal C}_{\rm ph}>1$ naturally leads to an effective phantom-crossing dark energy. Finally,  note that the phantom-crossing marker is related to the effective equation of state of dark energy, Eq.~\eqref{eq:w_eff_logrho}, by
\beq
\boxed{~{\cal C}_{\rm ph}(z) = 1 + \l(\f{1}{\Omega_m(z)}-1\r) \l[ 1 + w_{\rm DE}^{\rm eff}(z)\r] ~}\, .
\label{eq:Phantom_EoS_Diagnostic}
\eeq

\medskip

Before moving forward, we stress once more that effective phantom behaviour is fundamentally a statement about the inferred residual expansion history after subtracting an assumed pressureless matter contribution from the observed cosmological expansion. Consequently, different microscopic theories, including modified gravity, interacting dark sectors, or departures from the standard Friedmann equation, may all lead to similar reconstructed effective phantom dynamics even when none of the fundamental constituents individually satisfy Eq.~\eqref{eq:w_constituents}.  Next, we move on to briefly discuss the recent observational results~\cite{DESI:2024mwx,DESI:2024aqx,DESI:2025zgx,DESI:2025fii} and their appropriate interpretation within the reconstruction of dark energy. 

\section{Observational Hints of Dynamical Dark Energy}
\label{sec:DDE_DESI_Obs}
Before discussing the recent observational preference for dynamical dark energy, it is useful to briefly recall the observational setting in which such reconstructions arise. The Dark Energy Spectroscopic Instrument (DESI) measures BAO scales across a broad range of redshifts with unprecedented precision~\cite{DESI:2024mwx,DESI:2024aqx,DESI:2025zgx}. BAO observations constrain combinations of cosmological distances and the expansion history through quantities such as the angular diameter distance and Hubble parameter relative to the sound horizon scale. However, BAO measurements alone do not directly reconstruct the dark energy equation of state. In practice, the strongest constraints on dark energy emerge when DESI observations are combined with Cosmic Microwave Background (CMB) data~\cite{Planck:2018nkj,2020Planck,AtacamaCosmologyTelescope:2025blo,SPT-3G:2025bzu}, which calibrate the early Universe and the sound horizon scale, together with Type Ia supernova observations~\cite{Rubin:2023ovl,Brout:2022vxf,DES:2024tys,Popovic:2025glk,DES:2025sig}, which constrain the relative late time expansion history. The resulting analyses therefore reconstruct the effective dark energy sector within the framework discussed in Sec.~\ref{sec:Eff_DE}. In the following,  we briefly summarize the current observational situation in Sec.~\ref{sec:DESI_DR2}, and then clarify the appropriate interpretation of the recent DESI dynamical dark energy results in Sec.~\ref{sec:DESI_DR2_Implications}.

\subsection{Phantom-crossing Dark Energy in Light of Recent Observations}
\label{sec:DESI_DR2}
The recent DESI analyses have significantly sharpened observational constraints on the late time expansion history of the Universe. In particular, the combination of DESI BAO measurements with CMB and Type Ia supernova data has led to renewed interest in the possibility that the effective dark energy sector may be dynamical rather than a pure cosmological constant~\cite{DESI:2025zgx}. While the statistical significance of the present preference remains moderate and the interpretation continues to be actively debated~\cite{cortes2025desisdr2exclusionlambdacdm,Efstathiou:2025tie,Efstathiou:2024xcq,Herold:2025hkb,Ong:2025utx,Ong:2026tta,Dinda:2026ktu}, the resulting reconstructed evolution of the effective equation of state is nevertheless interesting from a phenomenological perspective~\cite{CosmoVerseNetwork:2025alb}.

Within the commonly used CPL parametrisation~\cite{Chevallier:2000qy,Linder:2002et},
\beq
w_{\rm DE}^{\rm eff} \equiv w_{\rm DE}^{\rm CPL}(z)=w_0+w_a \, \f{z}{1+z}\,,
\label{eq:wDE_CPL}
\eeq
the DESI DR2 analyses typically prefer regions of parameter space with
\beq
w_0>-1\,,\qquad w_a<0\,,
\label{eq:wDE_CPL_DESI}
\eeq
once DESI BAO data are combined with CMB and/or supernova observations. Representative best fit values range roughly in between~\cite{DESI:2025zgx}
\beq
w_0 \in \l(-0.9,\,-0.6\r)\,,\qquad w_a \in \l( -1.1,\,-0.4\r)\,,
\label{eq:wDE_CPL_DESI_Bestfit}
\eeq
 although the precise values vary somewhat between datasets and analysis choices. Such parameter values imply that at low enough redshifts
\beq
w_{\rm DE}^{\rm CPL}(z\ll 1) \simeq w_0 > -1\,,
\eeq
while at sufficiently high redshift,
\beq
w_{\rm DE}^{\rm CPL}(z\gg1)\simeq w_0+w_a<-1\,.
\eeq
Consequently, the reconstructed effective dark energy equation of state crosses the phantom divide during the recent cosmological past, corresponding precisely to the type of effective phantom crossing behaviour discussed earlier in Sec.~\ref{sec:DE_Phantom_Eff}.

In the CPL framework, the phantom crossing redshift is determined by demanding
$w(z_{\rm ph})=-1$ in Eq.~\eqref{eq:wDE_CPL}, which leads to
\beq
z_{\rm ph} = -\f{1+w_0}{1+w_0+w_a}\,,
\label{eq:CPL_z_phantom}
\eeq
which implies that the phantom crossing takes place in the past, $z_{\rm ph} > 0$, provided the right hand side is positive. For representative DESI like best fit values, this typically corresponds to $z_{\rm ph} \sim 0.5$.
Thus, the reconstructed effective dark energy behaves as phantom like with $w_{\rm DE}^{\rm eff} < -1$ for $z_{\rm ph} < z \leq z_{\rm max}$, while becoming non phantom, or quintessence like, with $w_{\rm DE}^{\rm eff} > -1$  for $z< z_{\rm ph}$ closer to the present epoch.
\begin{figure}[!t]
\begin{center}
\includegraphics[width=0.5\textwidth]{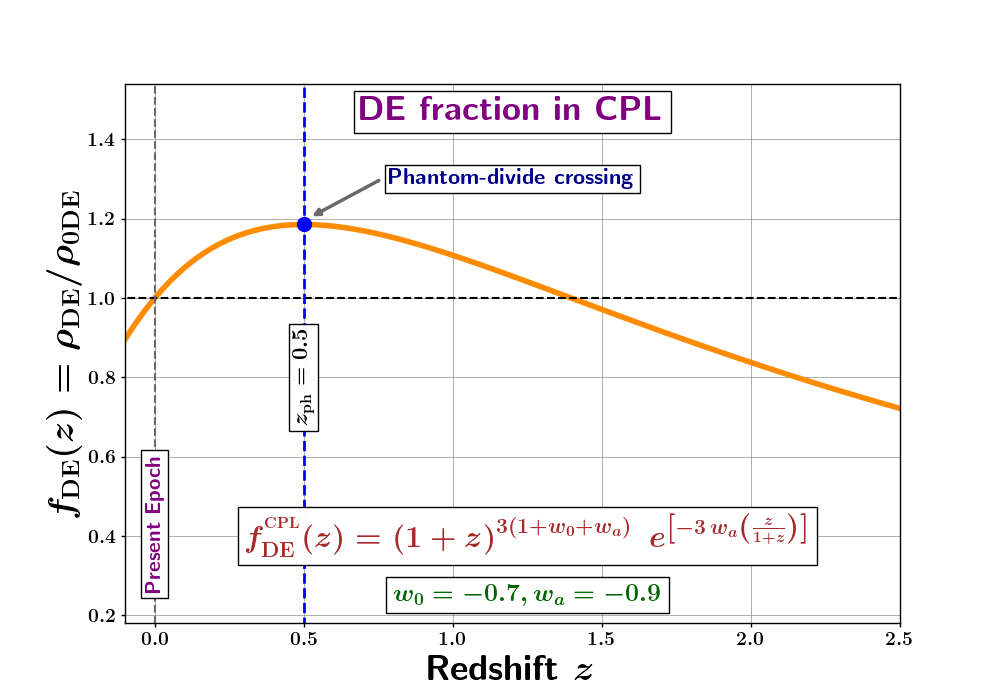}
\hspace{-0.2in}
\includegraphics[width=0.5\textwidth]{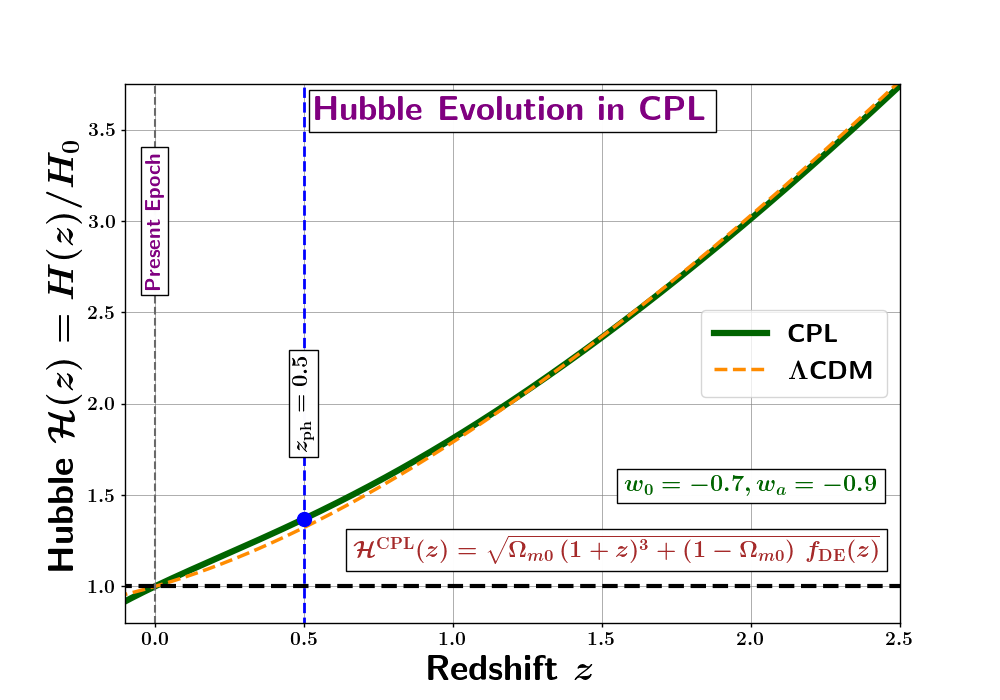}
\caption{{\bf Left panel} shows the evolution of DE density relative to its present-epoch value, defined in Eq.~\eqref{eq:fDE_CPL} for the effective EoS of dark energy in the CPL parametrisation, Eq.~\eqref{eq:wDE_CPL}, for  $\lbrace w_0 =-0.7,\,w_a=-0.9\rbrace$.  The dashed black line corresponds to the DE fraction in $\Lambda$CDM, namely, $f_{\rm DE}=1$. {\bf Right panel} shows the Hubble parameter corresponding to the phantom-crossing CPL reconstruction (solid green curve), compared with the $\Lambda$CDM expansion history (dashed orange curve), both  with $\Omega_{m0}=0.315$. The blue circle indicates the epoch of phantom-divide crossing, $w_{\rm DE}^{\rm eff}=-1$, identified from the left panel.}
\label{fig:DE_CPL}
\end{center}
\end{figure}

The corresponding reconstructed normalized effective dark energy density, as defined in Eq.~\eqref{eq:fde_def}, evolves as
\beq
f_{\rm DE}^{\rm CPL}(z)=(1+z)^{3\l(1+w_0+w_a\r)}\exp\l[-3\, w_a\,\f{z}{1+z}\r]\,.
\label{eq:fDE_CPL}
\eeq
As discussed in Sec.~\ref{sec:DE_Phantom_Eff}, phantom crossing corresponds to a change in the sign of the derivative of the reconstructed residual density. Consequently, the DESI preferred CPL region corresponds to an effective dark energy density that grows with cosmic time during the phantom phase for $z>z_{\rm ph}$, reaches a maximum near the crossing epoch $z=z_{\rm ph}$, and subsequently begins to decrease for $z<z_{\rm ph}$, once the reconstructed equation of state evolves back to $w_{\rm DE}^{\rm eff} > -1$. This is illustrated in Fig.~\ref{fig:DE_CPL}, for $w_0=-0.7,\,w_a=-0.9,~{\rm and}~\,\Omega_{m0}=0.315$. The solid orange curve in the left panel shows the evolution of $f_{\rm DE}^{\rm CPL}$, as given in Eq.~\eqref{eq:fDE_CPL}, while the solid green curve in the right panel displays the corresponding Hubble parameter (the dashed orange curve in the right panel corresponds to the Hubble parameter in $\Lambda$CDM). Blue circles correspond to the epoch of phantom-divide crossing.

Importantly, the {\em phantom behaviour is transient}, and the  EoS of dark energy at the present epoch itself is typically reconstructed to lie in the non-phantom regime. This point is conceptually important: the recent DESI preference is therefore not the standard scenario of eternally phantom dark energy with a constant equation of state satisfying $w_{\rm DE}^{\rm eff} < -1$
at all times. Rather, the preferred phenomenology corresponds to a temporary interval of effective phantom behaviour within the recent expansion history. In the language of Sec.~\ref{sec:DE_Phantom_Eff}, the reconstructed residual dark energy density grows with time over a finite redshift interval before beginning to decrease again at lower redshifts.

At present, the observational status of this behaviour, however,  remains unsettled. The statistical significance of the preference depends upon the precise combination of datasets, the treatment of systematics, the dark energy parametrisation adopted, and the statistical methodology used for model comparison~\cite{Jiang:2024xnu,cortes2025desisdr2exclusionlambdacdm,Efstathiou:2025tie,Efstathiou:2024xcq,Herold:2025hkb,Ong:2025utx,Ong:2026tta,Dinda:2026ktu}. In particular, discussions continue regarding possible observational systematics, tensions between different supernova compilations, prior dependence in Bayesian evidence calculations, and the extent to which the data genuinely require dynamical dark energy over $\Lambda$CDM. Moreover, while CPL provides a useful phenomenological description of broad smooth dark energy evolution\footnote{Note that the CPL parametrisation is phenomenologically motivated~\cite{Chevallier:2000qy,Linder:2002et}. Although it functionally resembles a low redshift expansion around the present epoch $a=a_0$, it was originally proposed primarily as a calibration of a broad class of dynamical dark energy models, see Ref.~\cite{Linder:2024rdj}. The author thanks Eric Linder for useful discussions and clarifications regarding this point.}, it is not designed to capture arbitrary rapid transitions, oscillatory behaviour, poles, or more complicated non local structures in the effective equation of state.

Nevertheless, an important aspect of the DESI extended dark energy analysis~\cite{DESI:2025fii} is that the qualitative preference for evolving dark energy does not appear to be tied exclusively to the CPL form itself. The DESI collaboration explored several parametric and non parametric reconstruction approaches and reported qualitatively similar trends favouring evolving effective dark energy, including scenarios featuring phantom crossing behaviour, although the significance varies between methods and no decisive conclusion can yet be drawn. Thus, while the present observational evidence for dynamical dark energy remains preliminary, the phenomenology motivating the present discussion is not restricted solely to a single parametrisation. We now move on to discuss the physical implications of these observational results in more detail.

\subsection{Implications of DESI results: What they Do and Do Not imply}
\label{sec:DESI_DR2_Implications}
The observational preference discussed in Sec.~\ref{sec:DESI_DR2} has generated considerable attention because phantom dark energy is often associated with microscopic ghost instabilities, violation of the null energy condition, and future cosmological singularities. However, as emphasized below, many of these conclusions do not immediately follow from the observational reconstruction of an effective dark energy sector. The recent DESI analyses reconstruct the background expansion history and infer the corresponding effective dark energy properties within the assumptions discussed earlier in Sec.~\ref{sec:Eff_DE_Assumptions}. Consequently, the physical interpretation of the reconstructed phantom and/or phantom-crossing behaviour, following the discussion in Sec.~\ref{sec:DE_Phantom_Eff},  requires some care. 

In the following, we discuss several important theoretical concerns commonly associated with phantom or phantom-crossing dark energy, and clarify why such conclusions do not automatically follow from the recent DESI reconstruction results.

\subsubsection*{\Brown Null Energy Condition:}
A particularly important distinction concerns the null energy condition (NEC) of a perfect fluid, given by $\rho_{\cal F} + p_{\cal F} \geq 0$, 
which is violated if the fluid satisfies $w_{\cal F}<-1$. Historically, this is one of the primary reasons why phantom dark energy attracted considerable theoretical concern. However, the observationally reconstructed quantity $w_{\rm DE}^{\rm eff}$ discussed in Sec.~\ref{sec:DE_Phantom_Eff} need not correspond directly to the equation of state of the underlying microscopic stress tensor itself. Instead, it characterizes the effective residual component inferred after rewriting the observed cosmological expansion history within the reconstruction framework discussed in Sec.~\ref{sec:Eff_DE_Assumptions}. Consequently, an effective phantom reconstruction does not by itself establish violation of the fundamental NEC by the underlying microscopic theory.

A related, but important point concerns the distinction between the effective dark energy equation of state, Eq.~\eqref{eq:w_q}, and the equation of state of the total cosmological fluid, Eq.~\eqref{eq:w_tot_def}. In FLRW cosmology, the NEC condition for the total cosmological fluid becomes
\beq
\boxed{~\bm{w_{\rm tot}(z) ~ \geq ~ -1 \quad \Longleftrightarrow \quad \rho_{\rm tot}(z) ~ + ~ p_{\rm tot}(z)  ~\geq ~ 0}~}\,,
\label{eq:GR_NEC}
\eeq

Consequently, even if the reconstructed effective dark energy temporarily satisfies $w_{\rm DE}^{\rm eff}(z)<-1$, the total cosmological fluid can still satisfy Eq.~\eqref{eq:GR_NEC}. Thus, transient effective phantom behaviour in the reconstructed dark energy sector does not by itself imply violation of the null energy condition by the total cosmological fluid. Indeed, for cosmologies close to the present DESI preferred region, the total equation of state typically remains comfortably above the NEC violating threshold throughout the observationally accessible cosmological history; for related discussions, see Ref.~\cite{Caldwell:2025inn}.

\subsubsection*{\Brown Big Rip Singularity:}
Similarly, the recent DESI preference for transient effective phantom behaviour does not automatically imply a future Big Rip singularity. The standard Big Rip scenario arises in cosmologies where a fundamentally phantom component with $w_{\cal F}<-1$ dominates the Universe eternally, leading to a monotonically growing dark energy density and divergent expansion within finite cosmic time. For a perfect fluid with constant $w_{\cal F}<-1$, the energy density evolves as
\beq
\rho_{\cal F}\propto a^{3\, \l|1+w_{\cal F}\r|}\,,
\eeq
which grows as the Universe expands, and can eventually drive a future finite time singularity~\cite{Caldwell:2003vq}. More general classes of future cosmological singularities have been systematically classified in Refs.~\cite{Nojiri:2005sx}. Moreover, even phantom like growth does not necessarily imply a standard Big Rip singularity, particularly for dynamically evolving equations of state, see Refs.~\cite{Gonzalez-Diaz:2003xmx,Frampton:2011rh,Frampton:2011sp,Astashenok:2012tv}.

\medskip

The crucial point in the context of DESI observations is that the preferred phenomenology corresponds only to {\em a transient interval of effective phantom behaviour} reconstructed over a finite redshift range. A finite interval of $w_{\rm DE}^{\rm eff}(z)<-1$ in the recent past is not sufficient to infer a future Big Rip singularity~\cite{Starobinsky:1999yw,Copeland:2006wr}. More precisely, a standard Big Rip requires the future cosmic proper time
\beq
\Delta t_{\rm Rip}\equiv t_{\rm Rip}-t_0= \int_{a_0}^{\infty}\f{\d a}{a\,H(a)}
\label{eq:future_time}
\eeq
to converge to a finite value. For the case of a dynamical (phantom) dark energy with an evolving EoS $w_{\rm DE}^{\rm eff}(a)$,  realising a Big Rip singularity in the finite future can be written, using Eqs.~\eqref{eq:Friedmann_GR_eff}~and~\eqref{eq:future_time}, as the condition that the quantity
\beq
\boxed{~\sqrt{\Omega_{m0}}\,H_0\, \Delta t_{\rm Rip} = \int_{1}^{\infty} \d s \l[\f{1}{s} + \l(\f{\Omega_{{\rm DE}0}}{\Omega_{m0}}\r)\,s^2\,\exp\l\{-3\int_{1}^{s} \d y \, \l( \f{1+w_{\rm DE}^{\rm eff}(y)}{y}\r)\r\} \r]^{-1/2}~}
\label{eq:BigRip_Condition}
\eeq
converges to a finite value. Eq.~\eqref{eq:BigRip_Condition} immediately demonstrates that the existence of a Big Rip singularity depends on the {\em future asymptotic behaviour} of the effective equation of state, rather than on whether $w_{\rm DE}^{\rm eff}(z)<-1$ over a finite interval in redshift. In particular, a transient epoch of effective phantom behaviour does not necessarily imply a future singularity. To see this, consider first the case of a constant phantom equation of state,
\beq
w_{\rm DE}^{\rm eff}(a) \equiv w < -1 \qquad (\text{Constant}) \, ,
\eeq
for which the exponential factor in Eq.~\eqref{eq:BigRip_Condition} then becomes
$$
\exp\l\{-3\int_1^s dy\,\f{1+w}{y}\r\} = s^{-3(1+w)} = s^{3|1+w|} \, .
$$
At sufficiently late times, when the dark energy contribution dominates over matter, we obtain 
$$
\Delta t_{\rm Rip} \propto \int^{\infty} \f{\d s}{s^{1+\f32|1+w|}} < \infty \, .
$$
which corresponds to the standard Big Rip singularity occurring at a finite future cosmic time. On the other hand, for a transient phantom epoch,  if the effective equation of state asymptotically approaches a cosmological constant,
\beq
w_{\rm DE}^{\rm eff}(a): ~~< -1 ~~\longrightarrow ~~ -1 \qquad \text{as} \qquad a \to \infty \, ,
\eeq
 sufficiently rapidly such that the integral $\int^\infty \d\ln y\,\l[1+w_{\rm DE}^{\rm eff}(y)\r]$ converges, then the exponential factor in Eq.~\eqref{eq:BigRip_Condition} tends to a constant, leading to 
$$
\Delta t_{\rm Rip} \propto \int^\infty \f{\d s}{s} ~\longrightarrow~ \infty \, .
$$
Consequently, {\em no Big Rip singularity occurs}. The Universe instead approaches an {\em asymptotic de Sitter} state~\cite{Sahni:2002dx,Astashenok:2012tv}.

Therefore, a finite interval of reconstructed phantom behaviour, such as that currently preferred by DESI analyses, is not by itself sufficient to infer a future Big Rip singularity. What matters is whether the effective phantom behaviour persists indefinitely into the future. In fact, if the reconstructed effective equation of state subsequently evolves back towards the non-phantom regime, the future lifetime of the Universe remains infinite despite the presence of a transient phantom phase. Thus, the current observational reconstruction does not imply the standard eternally phantom cosmology conventionally associated with a future Big Rip singularity.

\subsubsection*{\Brown Ghost Instability:} 
More generally, several physically well behaved cosmological scenarios can exhibit effective phantom or phantom crossing behaviour without introducing microscopic ghost instabilities or fundamental phantom fields. These include interacting dark sector models, scalar tensor theories and modified gravity, including braneworld cosmology. In such situations, the modified expansion history may be reinterpreted within standard GR as an effective dark energy component exhibiting phantom like behaviour even when none of the underlying microscopic degrees of freedom individually satisfy $w_{\cal F}<-1$. Consequently, effective phantom reconstruction should not be viewed as uniquely identifying a pathological microscopic theory, but rather as a possible phenomenological manifestation of a broader class of non standard cosmological dynamics~\cite{Ludwick:2017tox}.

The essential conceptual distinctions discussed in this work are summarized in Table~\ref{tab:Ph_Interpret} below, which emphasizes the distinction between reconstructed cosmological phenomenology  of dark energy and conclusions regarding the underlying microscopic theory. As stressed before, many of these conceptual distinctions are not fundamentally new, and several physically well behaved realizations of effective phantom or phantom crossing dark energy without microscopic pathologies were already discussed extensively in the literature during the early 2000s. However, the recent DESI results have once again highlighted the importance of clearly distinguishing between observationally reconstructed effective dark energy and the microscopic theory underlying cosmic acceleration. 

\begin{table}[htb]
\centering
\arrayrulecolor{tablegray}
\setlength{\arrayrulewidth}{1.8pt}
\renewcommand{\arraystretch}{2.5}
\setlength{\tabcolsep}{8pt}
\begin{tabular}{|p{0.28\textwidth}|p{0.3\textwidth}|p{0.3\textwidth}|}
\hline
\centering {\bf Phenomenology of Reconstructed DE} &
\centering {\bf Correct  Interpretation: Effective/Residual} &
\centering {\bf What it Does Not automatically Imply}
\tabularnewline
\hline
\centering Reconstructed EoS shows $w_{\rm DE}^{\rm eff}(z)<-1$ &
\centering Effective DE density $\rho_{\rm DE}^{\rm eff}(z)$ grows with cosmic time &
\centering Fundamental phantom field with wrong-sign kinetic term
\tabularnewline
\hline
\centering Effective phantom DE with~~ $\f{\d \rho_{\rm DE}^{\rm eff}}{\d t} > 0$ &
\centering With time, $H^2(z)$ falls slower than~~ $\rho_m(z) = \rho_{m0}\,(1+z)^3$ &
\centering Pathological Microscopic Theory or  Ghost Instability 
\tabularnewline
\hline
\centering Phantom-divide crossing  $w_{\rm DE}^{\rm eff}<-1 \, \longrightarrow  \, w_{\rm DE}^{\rm eff} > -1$ &
\centering Change in sign of $\f{\d \rho_{\rm DE}^{\rm eff}}{\d t}:~+\,{\rm ve}\,\longrightarrow\,-\,{\rm ve}$ &
\centering NEC Violation for the  total microscopic stress tensor
\tabularnewline
\hline
\centering Observational preference for transient phantom DE &
\centering Reconstructed $w_{\rm DE}^{\rm eff}(z)<-1$ over a finite redshift interval &
\centering An inevitable future Big Rip Singularity
\tabularnewline
\hline
\end{tabular}
\caption{Summary of the conceptual interpretation of observationally reconstructed effective phantom and phantom-crossing dark energy discussed in this work.}
\label{tab:Ph_Interpret}
\end{table}

With these conceptual distinctions clarified, we move on to briefly discuss several representative physical scenarios in which effective phantom behaviour may arise without  introducing fundamental phantom degrees of freedom in the next section.

\section{Physical realisations of Effective Phantom Dark Energy}
\label{sec:Eff_DE_Pheno}
As emphasized throughout this work, several physically well behaved cosmological scenarios exhibiting effective phantom dynamics have been extensively discussed in the literature since the early 2000s, motivated partly phenomenologically and partly by earlier dark energy reconstruction studies involving supernova observations~\cite{SupernovaSearchTeam:2004lze,Alam:2003fg,Riess:2006fw}. These include interacting dark sector models~\cite{Amendola:1999er,Farrar:2003uw,Hu:2004kh}, scalar-tensor and modified gravity theories~\cite{Esposito-Farese:2000pbo,Boisseau:2000pr,Bamba:2008hq,Clifton:2011jh,Joyce:2016vqv}, and extra-dimensional braneworld cosmology~\cite{Sahni:2002dx,Lombriser:2009xg,Bag:2018jle,Viznyuk:2018eiz,Alam:2016wpf,Bag:2016tvc}, metastable DE and vacuum decay~\cite{Parker:1999fc,Parker:2003as,Wang:2004cp,Li:2026hwq,Shafieloo:2016bpk,Li:2019san} and non-canonical kinetic sector~\cite{Wolf:2025acj,Nojiri:2026uvn}, to quote a few\footnote{Since the primary purpose of the present work is to clarify the physical interpretation of observationally reconstructed phantom and phantom-crossing dark energy, rather than to provide a review of  model building in light of the DESI results,  our discussion below focuses mainly on representative and historically important examples illustrating effective phantom behaviour and dark energy reconstruction.}. 

\bigskip

In all such cases, the effective phantom behaviour arises at the level of the reconstructed cosmological expansion history, even though the underlying microscopic theory itself need not contain any inherently phantom constituent satisfying $w_{\cal F}<-1$. In the following, we briefly mention a few  representative examples illustrating how such effective phantom behaviour can emerge.  We emphasize that the present discussion is restricted primarily to the reconstruction of dark energy at the homogeneous background level. A complete cosmological analysis of any specific model requires a proper treatment of cosmological perturbations, structure growth, and stability conditions, particularly when performing detailed parameter estimation using observational data, which we do not discuss in this work.

\subsection{Representative Scenarios in the Literature}
\label{sec:Models}
\subsubsection*{Interacting Dark Sector}
A particularly simple mechanism for generating effective phantom behaviour arises in interacting dark sector models~\cite{Amendola:1999er,Hu:2004kh,Wang:2016lxa}. As a simple illustrative  example, consider a scenario, where the underlying dark matter (dm) and dark energy (de) fluids exchange energy through an interaction term $Q$,
\beq
\dot{\rho}_{\rm dm} + 3\, H \, \rho_{\rm dm} = Q\,, \qquad \dot{\rho}_{\rm de} + 3\, H \, \l(1+w_{\rm de}\r)\rho_{\rm de} = -Q\,,
\label{eq:continuity_DMDE_int}
\eeq
so that the matter sector no longer evolves exactly as $\rho_{\rm dm} \propto (1+z)^3$. However, in accordance with the framework laid out in Sec.~\ref{sec:Eff_DE_Assumptions}, observational reconstruction of dark energy is  performed by assuming a conserved pressureless matter component,
\beq
\rho_m(z) \equiv \rho_{\rm CDM}(z) + \rho_{\rm b}(z)  = \rho_{m0}\,(1+z)^3\,,
\label{eq:rho_m_bCDM}
\eeq
where `b' denotes baryons, while `CDM' denotes cold dark matter, both being pressureless, with $\rho_{m0} = \rho_{{\rm CDM}0} + \rho_{{\rm b}0}$.  Consequently, rewriting the modified expansion history within the standard reconstruction framework of Sec.~\ref{sec:Eff_DE_Assumptions} using Eq.~\eqref{eq:rho_DE_eff_def} and after removing the conserved baryonic contribution, naturally leads to an effective dark energy density
\beq
\rho_{\rm DE}^{\rm eff}(z)= \rho_{\rm de}(z)+\rho_{\rm dm}(z)-\rho_{{\rm CDM}0}\,(1+z)^3\,,
\label{eq:rhoeff_DMDE_int}
\eeq
which can exhibit effective phantom or phantom-crossing behaviour even when the underlying dark sector constituents, $\lbrace {\rm dm},\,{\rm de}\rbrace$, themselves remain non-phantom. Recent discussions of interacting dark sector explanations of the DESI results may be found in Refs.~\cite{SantanaJunior:2024cug,Wang:2024vmw,vanderWesthuizen:2025vcb,vanderWesthuizen:2025mnw,vanderWesthuizen:2025rip,Shah:2025ayl,Abdalla:2026sis,Chakraborty:2025syu,Silva:2025hxw,Giare:2024smz,Andriot:2025los}.

\subsubsection*{Scalar-Tensor Theories and Modified Gravity}

Effective phantom behaviour can also arise naturally in scalar-tensor theories and modified gravity~\cite{Esposito-Farese:2000pbo,Boisseau:2000pr,Bamba:2008hq,Clifton:2011jh,Joyce:2016vqv}. In such scenarios, modifications of the gravitational sector alter the cosmological expansion history relative to standard GR. When the resulting background evolution is rewritten in the form of the standard Friedmann equation, these geometric modifications are absorbed into an effective dark energy sector. Consequently, the reconstructed effective equation of state can exhibit phantom or phantom-crossing behaviour even though the underlying theory itself does not contain a fundamental phantom field. Historically, scalar-tensor cosmology provided some of the earliest explicit examples of effective phantom acceleration without microscopic ghost instabilities~\cite{Esposito-Farese:2000pbo}. Similar effective phantom behaviour has subsequently been discussed extensively in other  modified gravity  based  cosmology~\cite{Amendola:2007nt,Clifton:2011jh,Joyce:2016vqv}. Recent discussions in light of the DESI results may be found in Refs.~\cite{Ye:2024ywg,Wolf:2024stt,Wolf:2025jed,Wolf:2025jlc,Pan:2025psn,Wang:2025znm,Wang:2026wrk,Koutroulis:2026gjr}.

\subsubsection*{Effective Phantom Behaviour from Braneworld Cosmology}
An  interesting and physically well motivated realisation of effective phantom dark energy arises in the ghost-free {\em normal branch}  braneworld scenario~\cite{Collins:2000yb,Dvali:2000hr,Shtanov:2000vr,Sahni:2002dx}. In such scenarios, the presence of a large non-compact  extra  dimension modifies gravitational dynamics on cosmological length scales, leading to departures from the standard Friedmann expansion history of GR. A particularly important consequence of these modifications is the natural emergence of effective phantom dark energy behaviour without introducing a fundamental phantom degree of freedom~\cite{Sahni:2002dx}. 

Indeed, braneworld cosmology provided one of the earliest explicit examples of stable effective phantom acceleration~\cite{Sahni:2002dx,Sahni:2005pf,Bag:2018jle,Viznyuk:2018eiz,Alam:2016wpf,Bag:2016tvc}. As a representative recent example, Ref.~\cite{Mishra:2025goj} demonstrated that braneworld cosmology can naturally realise effective phantom-crossing behaviour consistent with the DESI DR2 observations. Motivated by this, we briefly discuss the braneworld scenario below as an explicit illustration of the conceptual distinctions developed throughout this work.

\subsection{Crossing the phantom divide on the Braneworld}
\label{sec:Eff_DE_Braneworld}
In the braneworld scenario~\cite{ArkaniHamed:1998rs,Antoniadis:1998ig,Randall:1999vf,Randall:1999ee,Dvali:2000hr,Brax:2003fv,Brax:2004xh,Maartens:2010ar,Maartens:2003tw}, the  observable $(3+1)$-dimensional Universe is treated as a (mem)brane embedded in a higher-dimensional bulk spacetime. Standard (and beyond the Standard) Model fields are confined to the brane, whereas gravity propagates in both the brane and the bulk.  Cosmological dynamics on the {\em phantom braneworld}~\cite{Collins:2000yb,Shtanov:2000vr,Sahni:2002dx}   is governed by a simple  action of the form (following the normalisations used in Ref.~\cite{Mishra:2025goj})
\beq
S = M_{\rm p}^3 \int_{\rm bulk} {\cal R}_\text{B}  + \frac{\mpl^2}{2} \int_{\rm brane} \left( {\cal R} - 2 \, \Lambda_\sigma \right) + \int_{\rm brane} L_{\rm m} \, ,
\label{eq:Brane_Action}
\eeq
which includes the Einstein--Hilbert term both in the $(4+1)$-dimensional bulk (with Ricci curvature ${\cal R}_{\rm B}$) and on the $(3+1)$-dimensional brane (with Ricci curvature ${\cal R}$). Matter fields confined to the brane are described by the Lagrangian $L_{\rm m}$. The reduced Planck masses in the bulk and on the brane are denoted by $M_{\rm p}$ and $\mpl$, respectively, while  $\Lambda_\sigma$ denotes the brane tension (cosmological constant).  From the beginning, it is important to highlight that we work in the  normal branch of the braneworld which remains free from the catastrophic ghost instabilities~\cite{Gorbunov:2005zk, Koyama:2007za} that afflict the self-accelerating branch~\cite{Charmousis:2006pn}.

It is well known~\cite{Sahni:2002dx,Lue:2004za} that in the presence of a brane tension, $\Lambda_\sigma \neq 0$, the effective equation of state of dark energy is phantom like, $w_{\rm DE}^{\rm eff} < -1$. Following Ref.~\cite{Mishra:2025goj}  we  assume $\Lambda_\sigma = 0$, and attribute any effective non-zero vacuum energy at late times ($z \ll 100$) to a canonical scalar field. The expansion rate on a phantom brane   filled with pressureless matter, and a scalar field describing dynamical dark energy,  is then expressed as~\cite{Bag:2018jle,Mishra:2025goj} 
\beq
\boxed{~{\cal H} (z) = \sqrt{ \Omega_{m0}\, (1 + z)^3 + \frac{\rho_\phi}{3 m_p^2 H_0^2} + \Omega_{\ell 0}} - \sqrt{\Omega_{\ell 0}} ~}\, , 
\label{eq:H_BW}
\eeq
where 
\beq
\rho_\phi = \frac12 \dot \phi^2 + V (\phi) 
\label{eq:rho_phi}
\eeq
is the energy density of the scalar field. The effective braneworld parameter, $\Omega_{\ell 0}$, is given by 
\beq
\Omega_{0\ell} = \f{1}{\ell^2\,H_0^2}\, ; ~\quad~ \text{with} ~\quad~ \ell = \f{m_p^2}{M_p^3} \, ,
\label{eq:def_Omega_ell}
\eeq
where $\ell$ is the length scale above which effects of the extra spatial dimension become important~\cite{Sahni:2002dx}. Eq.~\eqref{eq:H_BW} demonstrates that if $\Omega_{\ell 0}$ is negligible (which is true at sufficiently high redshifts), then the braneworld evolution mimics that of the standard Friedmann evolution in GR. In fact, for $\Omega_{\ell 0} \to 0$, we recover the GR limit of braneworld gravity. 

\medskip

Effective density of dark energy in the braneworld, following the reconstruction framework of Sec.~\ref{sec:Eff_DE_Assumptions}, then becomes
\beq
3 m_p^2 H_0^2 {\cal H}^2(z) - \rho_m(z) \equiv \boxed{~\bm{\rho_{\rm DE}^{\rm eff}(z) = \rho_\phi(z) - 6 \, m_p^2 \,  H_0^2 \, \sqrt{\Omega_{\ell 0}}\, {\cal H}(z)}} \, .
\label{eq:DE_eff_BW} 
\eeq
This equation makes explicit the sense in which the braneworld modification appears as an effective dark energy component. From the perspective of a four-dimensional observer interpreting the background expansion history within standard GR, the extra-dimensional correction in Eq.~\eqref{eq:DE_eff_BW} is absorbed into the reconstructed effective dark energy density, and can therefore generate $w_{\rm DE}^{\rm eff}<-1$ without introducing any fundamental phantom degree of freedom. To see this, first consider the case where $\rho_\phi$ is replaced by the brane tension, $\Lambda_\sigma \neq 0$, or equivalently, if $\rho_\phi$ is a constant. Since the Universe is matter dominated at high enough redshifts, $z \gtrsim 1$, Hubble parameter decreases as ${\cal H} \propto (1+z)^{3/2}$. Consequently, given that $\rho_\phi$ is a constant, the effective dark energy density $\rho_{\rm DE}^{\rm eff}$ increases  at early times, displaying a phantom like dynamics, as originally noted in Ref.~\cite{Sahni:2002dx}. However, at late times, the matter density becomes negligible, leading to a nearly constant Hubble expansion, as can be inferred from  Eq.~\eqref{eq:H_BW}. Therefore, $\rho_{\rm DE}^{\rm eff}$ approaches a constant value, leading to an asymptotic de Sitter expansion without displaying a finite-future Big Rip singularity~\cite{Sahni:2002dx,Bag:2016tvc}, and without undergoing a phantom-crossing transition.  

\bigskip

In order to realise phantom-crossing dynamical dark energy on the braneworld, we now consider the case where $\rho_\phi$ is no longer constant, and the scalar field evolves down its potential $V(\phi)$ according to
\beq
\ddot \phi + 3 H \dot \phi + V_{,\phi}(\phi) = 0 \, ,
\label{eq:EoM_phi}
\eeq
where $V_{,\phi} \equiv {\d V}/{\d \phi}$. Canonical scalar field models are generally classified into two broad categories: {\em thawing} and {\em freezing} scalar fields~\cite{Caldwell:2005tm}. In thawing models, the scalar potential is sufficiently shallow that the slope term $V_{,\phi}$ initially remains subdominant compared to the Hubble friction term $3H\dot\phi$ in Eq.~\eqref{eq:EoM_phi}. Consequently, the scalar field remains nearly frozen at early times when $H \gg \sqrt{V_{,\phi\phi}}$, leading to an approximately constant energy density, $\rho_\phi \simeq {\rm constant}$. However, at late times, as the Hubble expansion rate drops below the effective mass scale of the scalar field, the field gradually begins to roll (thaw), resulting in a decaying dark energy component. As we discuss below, such thawing scalar fields play an important role in generating the effective phantom-crossing behaviour preferred by the DESI observations~\cite{Mishra:2025goj}. This behaviour contrasts with freezing scalar fields~\cite{Ratra:1987rm,Zlatev:1998tr,Bag:2017vjp}, whose cosmological evolution proceeds in the opposite direction~\cite{Bag:2018jle}.

\begin{figure}[!t]
\begin{center}
\includegraphics[width=0.5\textwidth]{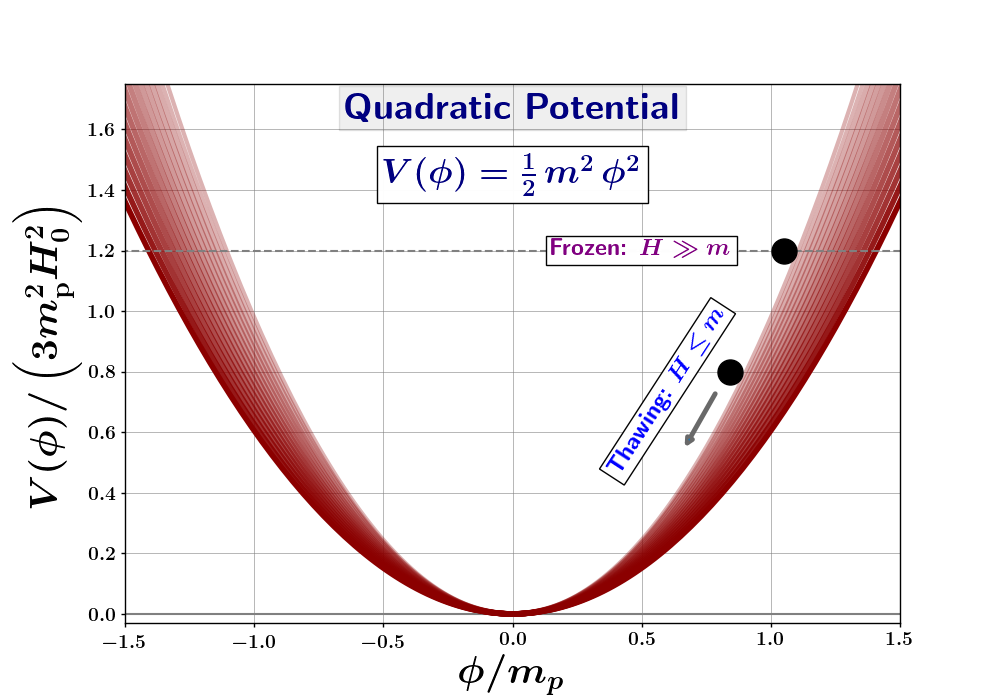}
\hspace{-0.2in}
\includegraphics[width=0.5\textwidth]{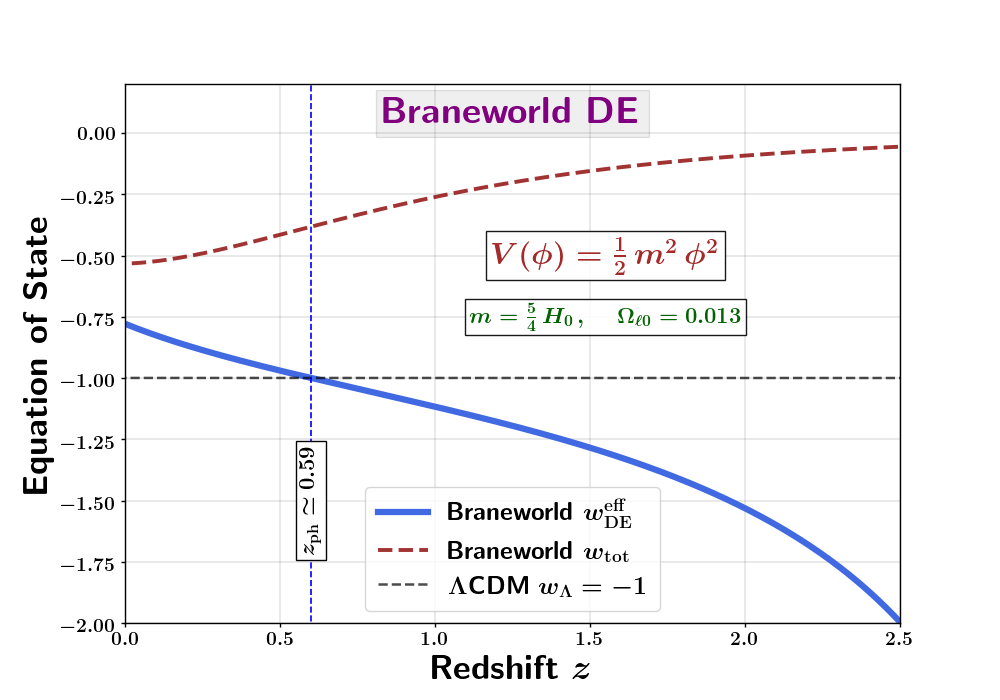}
\caption{{\bf Left panel} schematically shows the quadratic thawing potential, Eq.~\eqref{eq:pot_quad}, used to illustrate effective phantom-crossing behaviour on the braneworld. The scalar field remains frozen at early times when $H\gg m$, and begins to thaw and roll down the potential once $H\lesssim m$. {\bf Right panel} shows the corresponding reconstructed effective EoS of dark energy, Eq.~\eqref{eq:Brane_wDE_ext}, for the parameter choice in Eq.~\eqref{eq:best_fit_BW_quad}. The dashed horizontal line corresponds to $w_{\rm DE}^{\rm eff}=-1$, while the vertical blue line marks the phantom-divide crossing redshift $z_{\rm ph}\simeq0.59$. Note that the dashed red curve in the right panel shows the total effective EoS of the Universe $w_{\rm tot}$, defined in Eq.~\eqref{eq:w_tot_def}.}
\label{fig:DE_BW_pot_EoS}
\end{center}
\end{figure}

Since a thawing scalar field remains frozen at its initial value at early times, it effectively behaves like a cosmological constant (see the left panel of Fig.~\ref{fig:DE_BW_pot_EoS}). Consequently, following the discussion above, the effective dark energy density $\rho_{\rm DE}^{\rm eff}$ increases with time at higher redshifts, displaying a phantom like dynamics. However, at late times, as the scalar field begins to thaw and roll down its potential, both the terms $\rho_\phi$ and $6 \, m_p^2 \, H_0^2 \, \sqrt{\Omega_{\ell0}}\, {\cal H}(z)$ appearing on the right hand side of Eq.~\eqref{eq:DE_eff_BW} begin to decrease with cosmic time. If $\rho_\phi$ decreases sufficiently rapidly, then the effective dark energy density $\rho_{\rm DE}^{\rm eff}$ itself begins to decrease, thereby leading to a crossing of the phantom divide at some intermediate redshift.

Phantom-crossing dynamics on the braneworld can also be understood directly in terms of the effective equation of state of braneworld dark energy. Using Eq.~\eqref{eq:w_eff_logrho}, the effective equation of state can be written as~\cite{Mishra:2025goj}
\beq
w_\text{DE}^{\rm eff}(z)=-1-\l[\f{\sqrt{\Omega_{0\ell}}\,\Omega_{m}(z)-\f{1}{{\cal H}(z)} \,
\f{\dot{\phi}^2}{3m_p^2H_0^2}}{\bigl( 1- \Omega_{m}(z)\bigr)\l({\cal H}(z)+\sqrt{\Omega_{\ell0}}\r)}
\r] \, ,
\label{eq:Brane_wDE_ext}
\eeq
which demonstrates that at sufficiently high redshifts, when the scalar field remains frozen at its initial value, \textit{i.e.}, $\dot{\phi}^2 \simeq 0$, the effective dark energy equation of state is phantom like\footnote{Provided $\Omega_m \neq 1$, where $\Omega_m =1$ corresponds to an effective higher-redshift pole discussed in Refs.~\cite{Mishra:2025goj,Bag:2018jle,Sahni:2002dx}.}, satisfying $w_{\rm DE}^{\rm eff}(z)<-1$. At late times, however, as the kinetic contribution $\dot{\phi}^2$ becomes significant, the effective equation of state evolves back toward the non-phantom (quintessence-like) regime.

Although a wide variety of thawing scalar field potentials were shown in Ref.~\cite{Mishra:2025goj} to be compatible with the DESI observations, for simplicity we illustrate the above effective phantom-crossing behaviour using a quadratic potential of the form
\beq
V(\phi)=\f{1}{2}\,m^2\phi^2\, ,
\label{eq:pot_quad}
\eeq
which may be interpreted as a massive scalar field or an axion-like particle with a small mass $m\lesssim {\cal O}(H_0)$. Note that this is a two parameter extension of $\Lambda$CDM, namely, $\lbrace m,\,\Omega_{\ell 0}\rbrace$, which is similar to the case of CPL. The potential is schematically illustrated in the left panel of Fig.~\ref{fig:DE_BW_pot_EoS}.

\begin{figure}[!t]
\begin{center}
\includegraphics[width=0.75\textwidth]{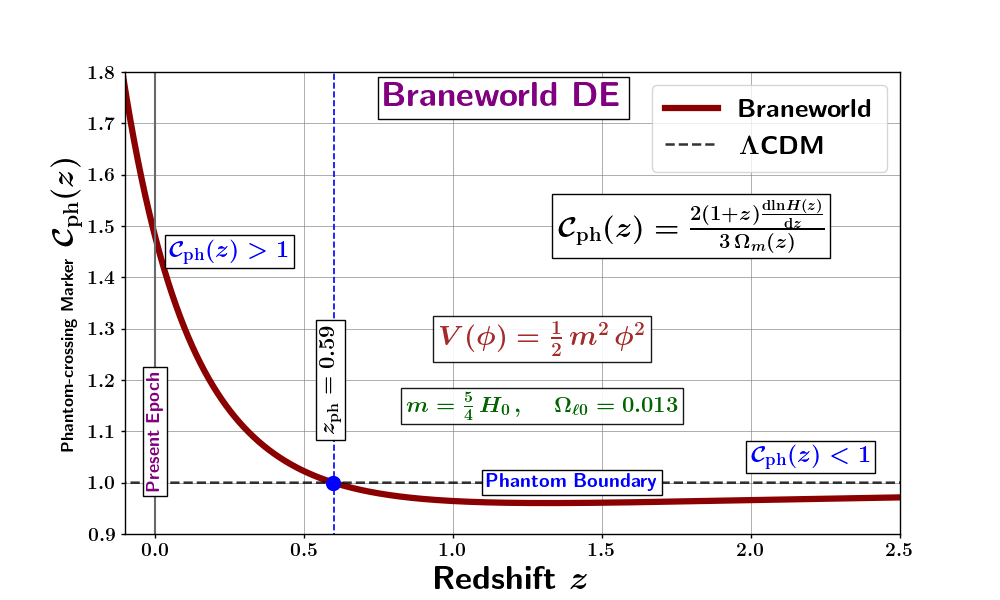}
\caption{Shows the phantom-crossing behaviour of the braneworld model with a quadratic thawing potential, Eq.~\eqref{eq:pot_quad}, in terms of the {\em phantom-crossing marker} ${\cal C}_{\rm ph}(z)$ defined in Eq.~\eqref{eq:Phantom_Diagnostic}. The dashed horizontal line corresponds to the phantom boundary, ${\cal C}_{\rm ph}=1$, separating the effective phantom regime, ${\cal C}_{\rm ph}<1$, from the non-phantom regime, ${\cal C}_{\rm ph}>1$. The vertical blue line marks the phantom-divide crossing redshift, $z_{\rm ph}\simeq 0.59$, at which ${\cal C}_{\rm ph}=1$.}
\label{fig:DE_BW_PhMarker}
\end{center}
\end{figure}

The corresponding evolution of the effective equation of state of dark energy, Eq.~\eqref{eq:Brane_wDE_ext}, is shown by the solid blue curve in the right panel of Fig.~\ref{fig:DE_BW_pot_EoS}. For comparison, the dashed red curve shows the total equation of state of the Universe, $w_{\rm tot}$, defined in Eq.~\eqref{eq:w_tot_def}. While the effective dark energy exhibits a phantom-divide crossing, the total equation of state remains above the phantom boundary, $w_{\rm tot} = -1$, indicating the total effective cosmological fluid remains above the NEC violating threshold. Consequently, despite the presence of an effective phantom phase, there is no indication of a finite-future Big Rip singularity. We also plot the {\em phantom-crossing marker}, ${\cal C}_{\rm ph}(z)$, defined in Eq.~\eqref{eq:Phantom_Diagnostic}, in Fig.~\ref{fig:DE_BW_PhMarker}. As expected, ${\cal C}_{\rm ph}(z)$ crosses the phantom boundary, ${\cal C}_{\rm ph}=1$, at $z_{\rm ph}\simeq 0.59$, in agreement with the phantom-divide crossing seen in the evolution of $w_{\rm DE}^{\rm eff}(z)$.

\begin{figure}[!t]
\vspace{-0.2in}
\begin{center}
\includegraphics[width=0.5\textwidth]{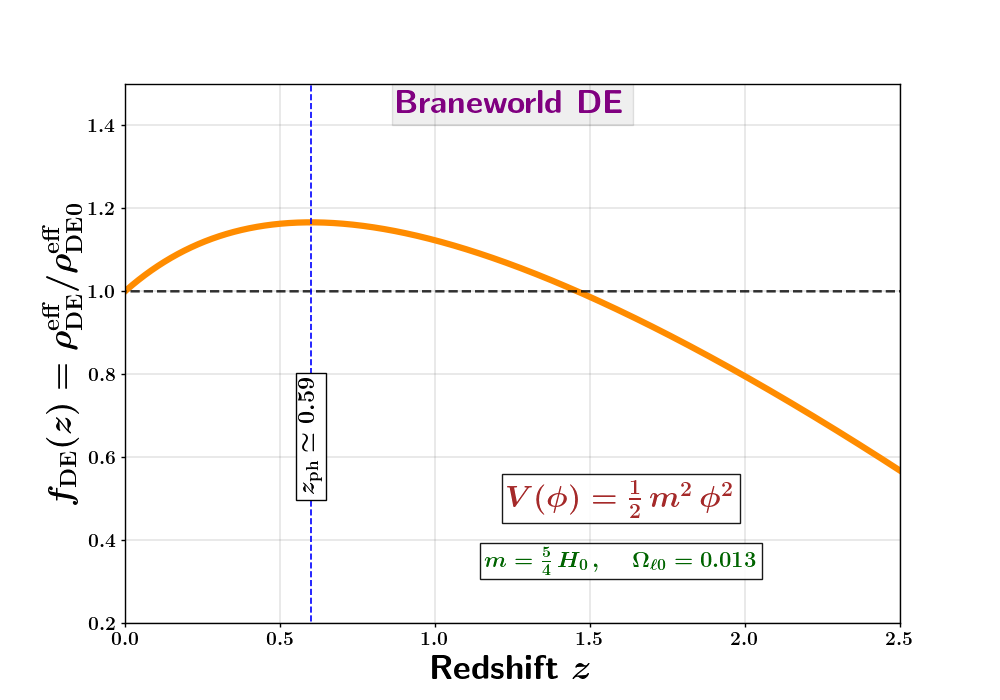}
\hspace{-0.2in}
\includegraphics[width=0.5\textwidth]{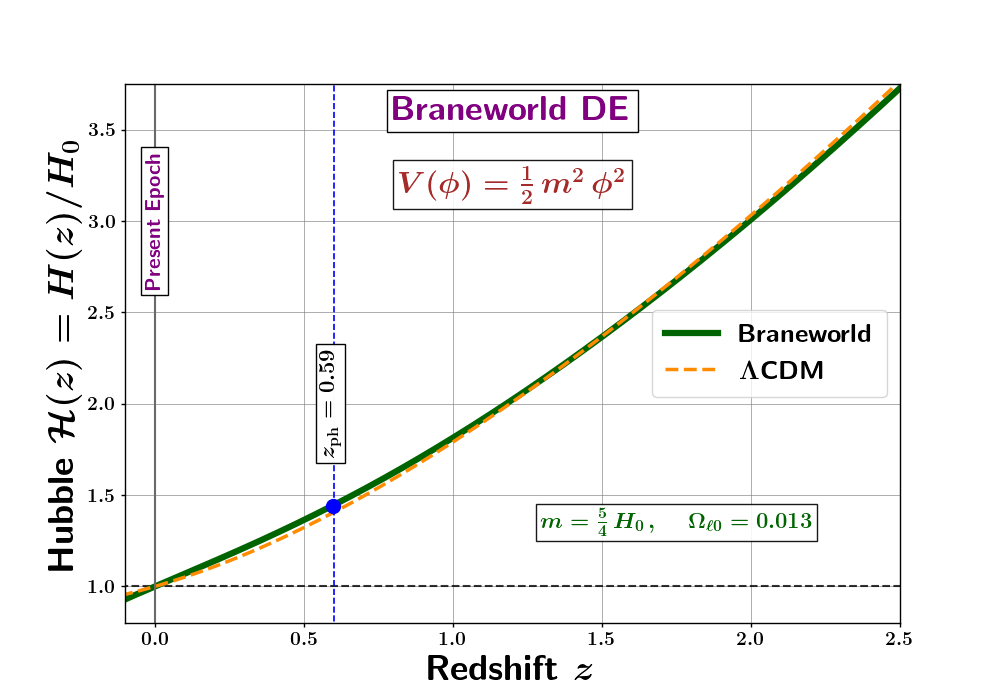}
\caption{{\bf Left panel} shows the evolution of the reconstructed effective DE density relative to its present-epoch value, $f_{\rm DE}(z)=\rho_{\rm DE}^{\rm eff}(z)/\rho_{{\rm DE}0}^{\rm eff}$, for the braneworld model with the quadratic thawing potential, Eq.~\eqref{eq:pot_quad}. The effective DE density grows with cosmic time during the phantom phase, reaches a maximum near the phantom-divide crossing redshift $z_{\rm ph}\simeq0.59$, and subsequently decreases at lower redshifts. {\bf Right panel} shows the Hubble parameter in the braneworld in solid green curve, while its value in $\Lambda$CDM is shown in dashed orange curve. The vertical blue line marks $z_{\rm ph}\simeq0.59$ in both panels.}
\label{fig:DE_BW_fDE_Hubble}
\end{center}
\end{figure}

Similarly, the evolution of the normalized effective dark energy density, $f_{\rm DE}(z)$, defined in Eq.~\eqref{eq:fde_def}, and the Hubble expansion rate, Eq.~\eqref{eq:H_BW}, are shown in the left and right panels of Fig.~\ref{fig:DE_BW_fDE_Hubble}, respectively. In the right panel, the braneworld Hubble parameter is shown by the solid green curve, while the dashed orange curve corresponds to its $\Lambda$CDM counterpart in GR. Consistent with the behaviour of the total equation of state, $w_{\rm tot}>-1$, shown by the dashed red curve in the right panel of Fig.~\ref{fig:DE_BW_pot_EoS}, the Hubble parameter decreases monotonically with cosmic time despite the presence of an effective phantom phase. This illustrates once again that effective phantom dark energy does not imply a phantom expansion of the Universe as a whole.

The braneworld quantities $w_{\rm DE}^{\rm eff}(z)$, ${\cal C}_{\rm ph}(z)$, $f_{\rm DE}(z)$ and ${\cal H}(z)$ are plotted in Figs.~\ref{fig:DE_BW_pot_EoS},\,\ref{fig:DE_BW_PhMarker}~and~\ref{fig:DE_BW_fDE_Hubble}, respectively, for the representative choice of parameters
\beq
m = \f{5}{4}\,H_0\,, \qquad \Omega_{\ell 0} = 0.013\, ,
\label{eq:best_fit_BW_quad}
\eeq
which lie close to the observationally best-fit region obtained in Ref.~\cite{Mishra:2025goj}. The value of $\Omega_{\ell0}$ is also compatible with existing bounds on the braneworld length scale from local gravity tests~\cite{Battat:2008bu,Dvali:2002vf,Koyama:2015vza}, as well as with constraints from the galaxy--ISW cross correlation~\cite{Lombriser:2009xg}, which suggest roughly $\Omega_{\ell 0}\lesssim 0.02$ in the corresponding phantom braneworld setup.

\bigskip

We emphasize, however, that the illustrative discussion here is restricted to background evolution. In the presence of a thawing scalar field and in light of the latest DESI results, a dedicated perturbation analysis and parameter estimation should be carried out separately. However, since the primary purpose of the present work is to illustrate how effective phantom and phantom-crossing dark energy can arise from physically well behaved models without any fundamental phantom degree of freedom, rather than to provide a complete explanation of the DESI observations, we do not discuss these issues further here and leave them to future work.

\section{Summary and Outlook}
\label{sec:Summary}
Recent DESI BAO measurements combined with CMB and supernova observations have renewed interest in the possibility of dynamical dark energy featuring phantom or phantom crossing behaviour. Motivated by the resulting discussions, and occasional conceptual confusion regarding the interpretation of such observational reconstructions, the primary goal of this paper has been to clarify:
\begin{itemize}
\item  What observationally reconstructed dark energy means in cosmology. 
\item  What conclusions may or may not be drawn from the recent observational preference for effective phantom behaviour.
\end{itemize}

We first reviewed the assumptions underlying the reconstruction of effective dark energy at the background level. In particular, observational reconstruction of dark energy relies upon: \textit{(i)}~the FLRW framework, \textit{(ii)}~the standard Friedmann equation of General Relativity, and \textit{(iii)}~a separately conserved pressureless matter sector at late times. Under these assumptions, any residual contribution required to explain the observed cosmological expansion history is interpreted as an effective dark energy component. Consequently, the reconstructed effective dark energy density and equation of state should be understood as effective quantities defined within a specific reconstruction framework, rather than direct probes of the microscopic theory underlying cosmic acceleration.

We then discussed the interpretation of effective phantom and phantom crossing behaviour. In particular, we emphasized that observationally reconstructed phantom behaviour corresponds fundamentally to a residual effective dark energy density that increases with cosmic time over part of the reconstructed expansion history. Similarly, phantom crossing corresponds to a change in the sign of the evolution of this reconstructed residual component. Importantly, these statements are first and foremost properties of the reconstructed background expansion history under the aforementioned specified assumptions. We introduced the {\em phantom-crossing marker}, a simple kinematic diagnostic constructed directly from the Hubble expansion history, which provides an intuitive characterization of effective phantom and phantom-crossing behaviour independent of the underlying microscopic realization.

We next discussed the recent DESI preference for dynamical dark energy and clarified several conclusions that do not automatically follow from such observational reconstructions. In particular, effective phantom behaviour does not by itself imply the existence of a fundamental phantom field featuring a  kinetic term with wrong sign, violation of the null energy condition by the total cosmological fluid, microscopic ghost instabilities, or an inevitable future Big Rip singularity. We also emphasized the distinction between the effective dark energy equation of state and the equation of state of the total cosmological fluid, which remains comfortably above the NEC violating threshold for cosmologies close to the present DESI preferred region.

More generally, physically well behaved realizations of effective phantom or phantom crossing dark energy without fundamental pathologies have been extensively discussed in the literature over the past two decades, including interacting dark sector scenarios, modified gravity, scalar tensor theories, and braneworld cosmology. Thus, many of the conceptual distinctions emphasized in this work are not fundamentally new. Rather, the purpose of the present work has been to restate these ideas explicitly and systematically in the context of the recent observational discussions surrounding dynamical dark energy.

An important extension of the present discussion would involve the inclusion of cosmological perturbations, where different microscopic realizations of effective phantom behaviour may be further distinguished through their predictions for structure growth, gravitational clustering, and modified gravitational dynamics.

\medskip

Finally, we stress that the present observational preference for dynamical dark energy remains statistically modest and its ultimate significance is still unclear. Future observations may strengthen, modify, or entirely remove the present indications for phantom crossing behaviour. Nevertheless, the conceptual distinctions discussed in this work are independent of the current observational status and remain important for correctly interpreting observationally reconstructed dark energy more generally. We therefore hope that this paper remains useful beyond the immediate context of the current observational results.

\section*{Acknowledgements}
SSM thanks the organisers and participants of {\em Dark Energy from Home} (DEfH\,2026) and {\em Gravity\,2026:~New Frontiers in Cosmology} international conferences for discussions. This manuscript owes its origin primarily to the latter conference where a preliminary version of the material was presented in the form of a talk. SSM is also thankful to Ed Copeland, Tony Padilla, Varun Sahni, Yuri Shtanov, Arman Shafieloo,  William Matthewson, Shadab Alam, Basudeb Dasgupta, Surhud More, and Suratna Das for exchanges on various aspects of the DESI results over the past one year. Thanks to Parth Bhargava and Sanket Dave for typo corrections and comments on the first draft.  SSM is supported by the Institute for Basic Science (IBS) as a Senior Researcher at CTPU-CGA under the project code, \texttt{IBS-R018-D3}.  \\

\noindent This work is purely theoretical and it does not involve any associated data. The generative AI tool \texttt{ChatGPT}$^+$ was used for limited stylistic and grammatical improvements of the manuscript.\\

\noindent For the purpose of open access, the author has applied a CC BY public copyright license to any Author Accepted Manuscript version arising. \\

\medskip

\begin{center}
\noindent{\Blue \bf \texttt{This work is dedicated to the loving memory of \\Prof.~P.~P.~Divakaran and Prof.~Naresh Dadhich.}}\\
\end{center}

\printbibliography

\end{document}